\documentclass[10pt,twocolumn,twoside]{IEEEtran} 
\usepackage{enumerate}
\usepackage{graphicx}
\usepackage{latexsym}
\usepackage{amsmath}
\usepackage{amssymb}
\usepackage{epsfig}
\usepackage{cite}
\usepackage{algorithmic}
\usepackage{overpic}
\usepackage{multirow}
\usepackage[table]{xcolor}
\usepackage{graphicx}
\usepackage{colortbl,array}
\usepackage{multirow,bigdelim}
\usepackage{graphicx}
\usepackage{subfigure}
\usepackage{booktabs}
\usepackage{booktabs,amsmath}
\usepackage{fancyhdr, graphicx, amsmath, amssymb}
\usepackage[linesnumbered,ruled]{algorithm2e}
\usepackage{arydshln}
\usepackage[english]{babel}
\usepackage{xcolor}
\usepackage{booktabs,amsmath}
\usepackage{fancyhdr, graphicx, amsmath, amssymb}
\usepackage{nomencl}
\newtheorem{thm}{Theorem}
\newtheorem{rem}{Remark}

\newtheorem{lem}{Lemma}
\newtheorem{prop}{Proposition}
\newtheorem{definition}{Definition}

\usepackage[linesnumbered,ruled]{algorithm2e}
\usepackage{arydshln}
\usepackage{cite}

\SetCommentSty{mycommfont}

\ifCLASSINFOpdf
\else
\fi

\begin{document}

\title{Strategic Topology Switching for Security--Part I: Consensus \& Switching Times}

\author{Yanbing~Mao, Emrah~Akyol,
        and Ziang~Zhang
\IEEEcompsocitemizethanks{\IEEEcompsocthanksitem Y.~Mao, E.~Akyol and Z.~Zhang are with the Department of Electrical and Computer Engineering, Binghamton University--SUNY, Binghamton, NY,
13902 USA, (e-mail: \{ymao3, eakyol, zhangzia\}@binghamton.edu). Parts of this paper were presented at the 57th IEEE Conference on Decision and Control, Miami Beach, FL, USA 2018 \cite{cdccs18}.}
\thanks{}}

\maketitle

\begin{abstract}
In this two-part paper, we consider strategic topology switching for the second-order multi-agent systems under a special class of stealthy attacks, namely the ``zero-dynamics" attack (ZDA). The main mathematical tool proposed here is to strategically switch the network topology to detect a possible ZDA. However, it is not clear a priori that such a switching strategy still yields consensus in this switched system, in the normal (un-attacked) operation mode.  In Part I, we propose a strategy on the switching times that enables the   topology-switching algorithm proposed in Part II~\cite{mao2017strategic} to reach the second-order consensus in the absence of a ZDA.  Utilizing the theory of stable switched linear systems with unstable subsystems,  we characterize sufficient conditions for the dwell time of topology-switching signal to reach consensus. Building on this characterization, we then propose  a decentralized time-dependent topology-switching algorithm.  The proposed  algorithm, used in conjunction with a simplified control protocol, achieves consensus while providing substantial advantages over other control approaches: it relies only on the relative position measurements (without any requirement for velocity measurements); and it does not impose any constraint on the magnitudes of coupling weights. We finally demonstrate our theoretical findings via the numerical simulation results.

\end{abstract}

\begin{IEEEkeywords}
Multi-agent systems, asymptotic second-order consensus, strategic topology switching, dwell time.
\end{IEEEkeywords}

\IEEEpeerreviewmaketitle

\section{Introduction}



\IEEEPARstart{T}{he} consensus of multi-agent systems with the first-order dynamics is a well-studied theoretical problem (see e.g., \cite{olfati2004consensus,jadbabaie2003coordination,fax2004information,ren2007distributed}) with many practical applications including decentralized computation~\cite{tsitsiklis1984problems}, distributed optimization~\cite{nedic2009distributed}, power sharing for droop-controlled inverters in islanded microgrids~\cite{lu2015consensus}, clock synchronization for sensor network~\cite{li2006global}, and more. However, current and emerging systems, such as vehicle~\cite{olfati2004consensus,ren2007distributed}, spacecraft~\cite{abdessameud2009attitude}, robot~\cite{chung2009cooperative} and
electrical power networks~\cite{johnson2014synchronization}, rely on the second-order dynamics. This observation, coupled with the fact that the consensus algorithms designed for the first-order multi-agent systems cannot be directly applied to those with the second-order dynamics, is the main motivation of this work.

A second-order multi-agent system consists of a population of $n$ agents whose dynamics are governed by the
following equations:
\begin{subequations}
\begin{align}
{\dot x_i}\left( t \right) &= {v_i}\left( t \right),\label{eq:oox1}\\
{{\dot v}_i}\left( t \right) &= u_{i}(t),\hspace{0.5cm}i = 1, \ldots ,n\label{eq:oox2}
\end{align}\label{eq:ooii}\end{subequations}
where $x_{i}(t) \in \mathbb{R}$ is the position, $v_{i}(t) \in \mathbb{R}$ is the velocity, and $u_{i}(t) \in \mathbb{R}$ is the control protocol of agent $i$.

Substantial research efforts have been devoted to the coordination control of the second-order multi-agent systems~(\ref{eq:ooii}), see e.g., the second-order consensus~\cite{ren2007distributed},  flocking~\cite{olfati2006flocking},  swarming~\cite{gazi2004stability},  velocity synchronization and regulation of relative distances~\cite{tanner2007flocking}, and with their applications in the decentralized formation control of mobile robots~\cite{lawton2003decentralized} and spacecrafts~\cite{ren2004decentralized}, and distributed continuous-time optimization~\cite{rahili2017distributed}, etc.
%

 In the following, we define the second-order consensus in this context.
\begin{definition}
\cite{yu2010some} The second-order consensus in the multi-agent system~(\ref{eq:ooii}) is achieved if and only if the following holds for any initial condition:  \begin{subequations}
\begin{align}
\mathop {\lim }\limits_{t \to \infty } | {{x_i}\left( t \right) - {x_j}\left( t \right)} | &= 0,\label{eq:oox1}\\
\mathop {\lim }\limits_{t \to \infty } | {{v_i}\left( t \right) - {v_j}\left( t \right)} | &= 0, \forall i,j = 1,
\ldots, n.\label{eq:oox2}
\end{align}\label{eq:ooid}\end{subequations}
\end{definition}

Conventional control protocols that achieve second-order consensus impose rather restrictive assumptions on the coupling strengths and network connectivity.  Recent research have focused on removing such restrictive assumptions. For example,  Mei et al.~\cite{mei2016distributed} propose an adaptive control gain to relax the conditions on the coupling strengths. Qin et al.~\cite{qin2016leaderless} derive lower bounds for coupling strengths of mechanisms of the leaderless and leader-following consensus. To deal with the problem of limited interaction ranges, Song and~You~\cite{ai2016second} propose a range-based varying weighs along coupling matrix that removes the assumption on the network connectivity.

\begin{table*}
\caption{Conditions of asymptotic second-order consensus under different control protocols.}
\label{msft}
\centering
\scalebox{0.80}[0.80]{%
\begin{tabular}{  p{1.25cm}  |  p{9.6cm} |  p{10.4cm}}
\bottomrule
\cellcolor[gray]{0.9} \textbf{Ref.} & \cellcolor[gray]{0.9} $ \cellcolor[gray]{0.9} \textbf{Control Protocol}$ &
\cellcolor[gray]{0.9} $\textbf{Constraints on}$\\  \hline
        \cite{yu2011second}   &${u_i}\left( t \right) = \alpha \sum\limits_{j = 1}^n {{{\tilde a}_{ij}}\left( {{x_j}\left( t \right) - {x_i}\left( t \right)} \right)}  - \beta \sum\limits_{j = 1}^n {{{\tilde a}_{ij}}\left( {{x_j}\left( {{t_k}} \right) - {x_i}\left( {{t_k}} \right)} \right)}$ & coupling strengths $\alpha$ and $\beta$, sampling period $t_{k+1}$ $-$ $t_{k}$, eigenvalues of Laplacian matrix\\ \hline
        \cite{huang2016some} & ${u_i}\left( t \right) = \alpha \sum\limits_{j = 1}^n {{{\tilde a}_{ij}}\left( {{x_j}\left( {{t_{k - 1}}} \right) - {x_i}\left( {{t_{k - 1}}} \right)} \right)}  - \beta \sum\limits_{j = 1}^n {{{\tilde a}_{ij}}\left( {{x_j}\left( {{t_k}} \right) - {x_i}\left( {{t_k}} \right)} \right)}$ &coupling strengths $\alpha$ and $\beta$, sampling period $t_{k+1}$ $-$ $t_{k}$, eigenvalues of Laplacian matrix \\ \hline
        \cite{abdessameud2010consensus}  & $\left\{ \begin{array}{l}
\!\!\!\!{u_i}\left( t \right) =  - \sum\limits_{j = 1}^n {{k_{ij}}\tanh \left( {{\lambda ^r}\left( {{x_j}\left( t \right) - {x_i}\left( t \right)} \right)} \right)}  + {{\dot \phi }_i}\left( t \right)\\
\!\!\!\!{{\dot \phi }_i}\left( t \right) =  - k_i^\phi \tanh \left( {{\lambda ^\phi }{\phi _i}\left( t \right)} \right) - \sum\limits_{j = 1}^n {{k_{ij}}\tanh \left( {{\lambda ^r}\left( {{x_j}\left( t \right) - {x_i}\left( t \right)} \right)} \right)}
\end{array} \right.$  & scalar gains ${\lambda ^\phi }$ and ${\lambda ^r}$, coupling weights $k_{ij}$, bound of distributed control inputs $u_{i}(t)$\\\hline
        \cite{xia2018event}  & ${u_i}\left( t \right) = \alpha \sum\limits_{j = 1}^n {{{\tilde a}_{ij}}( {{x_j}({t_{k^{'}_j\left( t \right)}^jh}) - {x_i}({t_k^ih})})}  - \beta \sum\limits_{j = 1}^n {{{\tilde a}_{ij}}({{x_j}({t_{k^{'}_j}^jh}) - {x_i}({t_{k - 1}^ih})})}$  & coupling strengths $\alpha$ and $\beta$, sampling period $h$, eigenvalues of Laplacian matrix\\ \hline
        \cite{Song17}  & $\left\{ \begin{array}{l}
{u_i}\left( t \right) =  - \beta \sum\limits_{j = 1}^n {{a_{ij}}\left( {{x_i}\left( {{t_k}} \right) - {x_j}\left( {{t_k}} \right)} \right)}  - {w_i}\left( t \right)\\
{{\dot w}_i}\left( t \right) =  - \alpha \sum\limits_{j = 1}^n {{a_{ij}}\left( {{x_i}\left( {{t_k}} \right) - {x_j}\left( {{t_k}} \right)} \right)}  - {w_i}\left( t \right)
\end{array} \right.$  & coupling strengths $\alpha$ and $\beta$, sampling period $t_{k+1} - t_{k}$, eigenvalues of Laplacian matrix\\
        \bottomrule
\end{tabular}
}
\end{table*}

However, the majority of the aforementioned prior approaches require the measurements of relative positions as well as the individual/relative velocities~\cite{ren2007distributed,ai2016second,mei2016distributed,yu2010some,qin2016leaderless}. A few recent control protocols that require only relative position measurements are summarized in Table I. 
In this paper, we provide a systematic study on whether the control protocols summarized in Table I can be simplified to
\begin{align}
u_{i}(t) = \sum\limits_{j = 1}^n {a_{ij}\left({{x_j}\left( t \right) -  {x_i}\left( t \right)} \right)}, i = 1, \ldots, n.\label{eq:ooxhh}
\end{align}

%

We note in passing that in prior work,  topology changes have been conventionally treated as disturbances that needs to be mitigated~\cite{jadbabaie2003coordination,ren2005consensus,psillakis2017consensus,lin2010consensus,su2016distributed}. To the best of our knowledge, active/strategic topology switching for networked control systems has not been systemically studied, with exceptions being~\cite{xie2005consensus,M18,mao2018synchronization,allerton18}.  Xie and Wang~\cite{xie2005consensus} proposed a centralized topology-switching algorithm to achieve consensus for the first-order dynamics. 
We note that Mao and Zhang proposed a decentralized  state-dependent topology-switching algorithm that achieves the second-order consensus~\cite{M18}. However, the  approach in ~\cite{M18} rules out only the Chattering Zeno behavior (zero dwell time of switching topologies~\cite{1582237, 1470118}), i.e., it might have Genuinely Zeno behavior (nonzero dwell time but infinitely switching over finite time~\cite{1582237, 1470118}). The approach taken in this paper eliminates both of these undesired behaviors. 

Stealthy attacks, particularly the ``zero-dynamics" attack (ZDA), poses an existential threat to wide deployment of networked control systems. ZDA essentially hides its attack signal in the null-space of the dynamics, and hence it cannot be detected via conventional detection methods, i.e., it has "zero" impact on the ``dynamics". Recent experiments ~\cite{teixeira2012attack} demonstrate that changing the system dynamics can be a remedy for ZDA.  Motivated by this observation, we consider topology switching as an effective way of altering the system dynamics and hence detecting ZDA. However, an important concern is the following: Can switching the topology destroy the  stability in the absence of attacks (in normal operation mode)?   The strategy on switching times proposed in this Part-I paper addresses this problem. Another relevant question that pertains to the design of topologies that can be used to detect ZDA  is studied explicitly in Part-II of this two-part paper~\cite{mao2017strategic}.


The contribution of this paper is twofold, which can be summarized as follows:
\begin{itemize}
  \item We propose a simplified second-order control protocol that only requires measurements of relative positions. Building on the well-known results on the stability of switched linear systems with unstable subsystems, we obtain a strategy on the dwell time of topology-switching signal that guarantees consensus in the absence of attacks.
  \item We propose a decentralized time-dependent topology-switching algorithm, based on the derived characterization of switching times. The proposed algorithm achieves second-order consensus without imposing any constraints on the magnitudes of coupling weights, in sharp contrast with prior work. 
\end{itemize}


This paper is organized as follows. In Sections II and III we present the notation and Problem formulation respectively. In Section IV, we derive the strategy on switching times. We present a decentralized time-dependent topology-switching algorithm in  Section V. Numerical examples are provided in Section VI. Finally, we discuss our conclusions in Section VII.

\section{NOTATION}
 We use $P > 0$ $(\geq, <, \leq 0)$ to denote a positive definite (positive semi-definite, negative
definite, negative semi-definite) matrix $P$. We let $\mathbb{R}^{n}$ and $\mathbb{R}^{m \times n}$ denote the sets of $\emph{n}$-dimensional real vectors and $m \times n$-dimensional real matrices, respectively. The symbol $\mathbb{N}$ represents the set of natural numbers and $\mathbb{N}_{0} = \mathbb{N} \cup \{0\}$. We let $\mathbf{I}$ and $\mathbf{0}$ denote the identity matrix and the zero matrix with compatible dimension, and  $\mathbf{1}_{n} \in \mathbb{R}^{n}$ and $\mathbf{0}_{n} \in \mathbb{R}^{n}$ denote the vector with all ones and the vector with all zeros, respectively. The superscript `$\top$' stands for matrix transpose. $\mathcal{K}_{\infty}$ denotes the set of strictly increasing, continuous and unbounded functions $[0, \infty) \rightarrow [0, \infty)$ which is zero at zero. For a symmetric matrix $M \in \mathbb{R}^{n \times n}$, we arrange its eigenvalues in an increasing order as $\lambda_{1}\left( M \right) \le \lambda_{2}\left( M \right) \le \ldots \le \lambda_{n}\left( M \right)$. $\mathfrak{S}$ is the set of indices of switching topologies (or the set of indices of subsystems of switched systems). $\emph{\emph{lcm}}(\cdot)$ denotes the operator of least common multiple among scalers. $\mathbb{Q}$ stands for the set of rational numbers.

The interaction among $n$ agents is modeled by an undirected graph $\mathrm{G} = (\mathbb{V}, \mathbb{E})$, where
$\mathbb{V}$ = $\left\{ {1,2, \ldots, n} \right\}$ is the set of vertices that represent $n$ agents and
$\mathbb{E} \subset \mathbb{V} \times \mathbb{V}$ is the set of edges of the graph $\mathrm{G}$. The weighted adjacency matrix $\mathcal A = \left[ {{a_{ij}}} \right]$ $\in \mathbb{R}^{n \times n}$ of the undirected graph $\mathrm{G}$ is defined as $a_{ij} = a_{ji} > 0$ if $(i, j) \in \mathbb{E}$, and $a_{ij} = a_{ji} = 0$ otherwise. Moreover, by convention, the undirected graphs do not have self-loops, i.e., for any ${i} \in \mathbb{V}$, $a_{ii} = 0$.  The Laplacian matrix of a graph $\mathrm{G}$ is defined as $\mathcal{L} = \left[ {{l_{ij}}} \right] \in {\mathbb{R}^{n \times n}}$, where ${l_{ii}} = \sum\limits_{j = 1}^n {{a_{ij}}}$ and ${l_{ij}} =  - {a_{ij}}$ for $i \neq j$. The diameter $d$ of a graph is the longest shortest un-weighted path between any two vertices in the graph.

\section{Problem Formulation}

The second-order multi-agent system~(\ref{eq:ooii}) under the simplified control protocol~(\ref{eq:ooxhh}) that involves topology switching is described by
\begin{subequations}
\begin{align}
\!\!\!\!{\dot x_i}\left( t \right) & = {v_i}\left( t \right)\label{eq:oon1}\\
\!\!\!\!{{\dot v}_i}\left( t \right) & = \!\sum\limits_{j = 1}^n {a_{ij}^{\sigma \left( t \right)}\left(
{{x_j}\left( t \right) -  {x_i}\left( t \right)} \right)}, i = 1, \ldots ,n\label{eq:oon2}
\end{align}\label{eq:oon}\end{subequations}
where
\begin{itemize}
  \item $\sigma (t):[t_{0},\infty ) \to \mathfrak{S} \triangleq \{1,2,\ldots,s\}$, is the switching signal of the interaction topology of the communication network, i.e., $\sigma (t) = {p_k} \in \mathfrak{S}$ for $t \in [t_{k}, t_{k+1})$ means the $p^{\emph{\emph{th}}}$ topology is activated over the time interval $[t_{k}, t_{k+1}), k \in \mathbb{N}_{0}$;
  \item $a^{p_{k}}_{ij}$ is the entry of the weighted adjacency matrix that describes the activated $p^{\emph{\emph{th}}}$ topology over the time interval $[t_{k}, t_{k+1}), k \in \mathbb{N}_{0}$.
\end{itemize}
We refer to $a^{\sigma(t)}_{ij} \ge 0$  as the \textit{coupling weights.}  For an undirected topology, $a^{\sigma(t)}_{ij} = a^{\sigma(t)}_{ji}$, it is straightforward to verify from~(\ref{eq:oon2}) that $\sum\limits_{i = 1}^n {{\dot{v}_i}\left( t \right)}  = 0$, for $t \ge 0$, which implies that the average position \begin{equation}
\bar x(t) \triangleq \frac{1}{n}\sum\limits_{i = 1}^n {{x_i}\left( 0 \right)}  + \frac{1}{n}\sum\limits_{i = 1}^n {{v_i}\left(0 \right)} t,\label{eq:das}
\end{equation}
proceeds with the constant velocity
\begin{equation}
\bar{v} \triangleq \frac{1}{n}\sum\limits_{i = 1}^n {{v_i}(t)}  = \frac{1}{n}\sum\limits_{i = 1}^n
{{v_i}({0})}.\label{eq:dav}
\end{equation}

If the second-order consensus is achieved, the individual velocities will converge asymptotically to the average of initial of
velocities, i.e., $\mathop {\lim }\limits_{t \to \infty } \left| {{v_i}\left( t \right) - \bar v} \right| = 0,i = 1,
\ldots ,n$. Based on relations~(\ref{eq:das}) and~(\ref{eq:dav}), we define the following fluctuation terms:
\begin{subequations}
\begin{align}
{{{\tilde x}}_i}\left( t \right) &\triangleq {x_i}\left( t \right) - \bar{x}(t),\label{eq:flu1}\\
{{\tilde v}_i}\left( t \right) &\triangleq {v_i}\left( t \right) - \bar v.\label{eq:flu2}
\end{align}\label{eq:fluv}
\end{subequations}
The dynamics~(\ref{eq:oon}) can  be expressed equivalently as
\begin{subequations}
\begin{align}
{\dot{\tilde{x}}_i}\left( t \right) &= {\tilde{v}_i}\left( t \right) \label{eq:foon1}\\
{{\dot{\tilde{v}}}_i}\left( t \right) &= \sum\limits_{j = 1}^n {a_{ij}^{\sigma \left( t
\right)}\left( {{\tilde{x}_j}\left( t \right) - {\tilde{x}_i}\left( t \right)} \right)}, i = 1, \ldots
,n.\label{eq:foon2}
\end{align}\label{eq:foon}\end{subequations}

It follows from~(\ref{eq:das})--(\ref{eq:fluv}) that
\begin{align}
{\bf{1}}_n^ \top \tilde x \left( t \right) &= 0, \hspace{0.1cm} \emph{\emph{for}} \hspace{0.1cm} t \geq 0 \label{eq:usef1}\\
{\bf{1}}_n^ \top \tilde v \left( t \right) &= 0, \hspace{0.1cm} \emph{\emph{for}} \hspace{0.1cm} t \geq 0. \label{eq:usef1no}
\end{align}

We next present the conditions on topology set $\mathfrak{S}$ for consensus in this part-I paper, and observer design in part-II paper~\cite{mao2017strategic}.
\begin{subequations}
\begin{align}
&\forall r \in \mathfrak{S}: \sqrt {\frac{{{\lambda _i}\left( {{{\cal L}_r}} \right)}}{{{\lambda _j}\left( {{{\cal L}_r}} \right)}}} \in \mathbb{Q}, \hspace{0.1cm} \emph{\emph{for}} \hspace{0.1cm} \forall i,j = 2, \ldots ,n\label{eq:bss1}\\
&\exists r \in \mathfrak{S}: \mathcal{L}_{r} \hspace{0.1cm} \emph{\emph{has}} \hspace{0.1cm} \emph{\emph{distinct}} \hspace{0.1cm} \emph{\emph{eigenvalues}}.\label{eq:bss2}
\end{align}\label{eq:bss}
\end{subequations}

In Section IV, we first show that the multi-agent system~(\ref{eq:foon}) under fixed topology, i.e, $\sigma(t) = p \in \mathfrak{S}$ for $t \in [0, \infty)$, is oscillating. This implies that using the simplified control protocol~(\ref{eq:oon2}) \emph{under fixed topology}, the multi-agent system~(\ref{eq:oon}) cannot achieve the second-order consensus, even for large coupling weights $a^{\sigma(t)}_{ij}$. Fortunately, using the stability of switched linear system (Lemma~\ref{thm:sst}) and the period of the system~(\ref{eq:foon}) under fixed topology, a strategy on switching times is derived, which enables the control protocol in~(\ref{eq:oon2}) to achieve consensus. 

\section{Strategy on Switching Times}
We next present


\begin{lem}
Consider the following system
\begin{equation}
\dot{\tilde x}\left( t \right) =  - \mathcal{L}_{r}\int_0^t {\tilde x\left( \tau  \right)} \mathrm{d}\tau +
\tilde{v}(0), t \geq 0\label{eq:add}
\end{equation}
where $\mathcal{L}_{r} \in \mathbb{R}^{n \times n}$ is the Laplacian matrix of a connected undirected
graph; and ${\tilde{x}}\left( t \right)\in {\mathbb{R}^n}$ and ${\tilde{v}}\left( t \right) = {\dot{\tilde{x}}}\left( t \right) \in {\mathbb{R}^n}$
satisfy~(\ref{eq:usef1}) and~(\ref{eq:usef1no}), respectively. The solutions of $\tilde{x}_{i}(t), i = 1, \ldots, n$, are
\begin{align}
&{{\tilde x}_i}\left( t \right)  \label{eq:LP12ka} \\
& = \!\!\sum\limits_{l = 2}^n {{q_{li}}q_{l}^{\top}\!\!\left(\!\!{\tilde x\!\left( 0 \right)\!\cos\! \left(\!t\sqrt {{\lambda_l}(\mathcal{L}_{r})} \right) \!+\!
\frac{{\tilde v\left( 0 \right)}}{{\sqrt {{\lambda_l}(\mathcal{L}_{r})} }}\!\sin\!\left(\!t\sqrt {{\lambda_l}(\mathcal{L}_{r})} \right)} \!\!\right)} \nonumber
\end{align}
where ${q_l} = \left[ {{q_{l1}}, \ldots ,{q_{ln}}} \right]^{\top} \in {\mathbb{R}^n}$ are the orthogonal vectors associated with eigenvalues ${\lambda_l}(\mathcal{L}_{r})$ (${\lambda_1}(\mathcal{L}_{r})$ = 0), $l = 2, \ldots, n$.
\label{thm:my0}
\end{lem}
\begin{IEEEproof}
See Appendix A.
\end{IEEEproof}

The system under switching topology~(\ref{eq:foon}) can be viewed as a switched linear system~(\ref{eq:sw}). Its equilibrium point is $\left( {{{\tilde x}^*},{{\tilde v}^*}} \right) = \left( {{\mathbf{0}_n},{\mathbf{0}_n}} \right)$. Let ${\sigma}(t) = r$ $\in$ $\mathfrak{S}$ for $t \in \left[ {{t_k},t_{k + 1}} \right)$, $k \in \mathbb{N}_{0}$, the subsystem of~(\ref{eq:foon}) can be rewritten as $\dot{\tilde x}\left( t \right)$ $=$  $- {{\cal L}_r}\int_{{t_k}}^t {\tilde x\left( \tau  \right)} {\rm{d}}\tau$ $+$ $\tilde v({t_k}),t \in \left[ {{t_k},{t_{k + 1}}} \right)$. Then, Lemma~\ref{thm:my0} implies that each subsystem of the switched system~(\ref{eq:foon}), i.e., the multi-agent system~(\ref{eq:foon}) under each fixed topology, is not stable. Hence, the problem of strategic topology switching would be designing the stabilizing switching rule for the switched systems with unstable subsystems. Let us first recall a technical lemma regarding the stability of switched linear systems without stable subsystems, which will be used to derive a strategy on switching times for consensus.
\begin{lem}[Stability of Switched Systems without Stable Subsystems~\cite{xiang2014stabilization}]
Consider a switched linear system
\begin{equation}
\dot z\left( t \right) = {A_{\sigma (t)}}z\left( t \right),\label{eq:sw}
\end{equation}
where $z\left( t \right) \in {\mathbb{R}^m}$, ${A_{\sigma (t)}} \in {\mathbb{R}^{m \times m}}$ and $\sigma(t) \in \mathfrak{S}$. Given scalars $\alpha > 0$, $1> \beta > 0$, $ \widehat{\tau}_{\max} \geq \widehat{\tau}_{\min} >0$ and $\kappa \in \mathbb{N}$, if there exists a set of matrices $P_{r,q} > 0, q = 0, 1, \ldots, \kappa$, $r \in {\mathfrak{S}}$, such that $\forall q = 0, 1, \ldots, \kappa-1$, $\forall r, s \in {\mathfrak{S}}$, such that
\begin{align}
&A_r^\top{P_{r,q}} + {P_{r,q}}{A_r} + \Psi _r^q - \alpha {P_{r,q}} < 0,\label{eq:lm1}\\
&A_r^\top{P_{r,q + 1}} + {P_{r,q + 1}}{A_r} + \Psi _r^q - \alpha {P_{r,q + 1}} < 0,\label{eq:lm2}\\
&A_r^\top{P_{r,\kappa}} + {P_{r,\kappa}}{A_r} - \alpha {P_{r,\kappa}} < 0,\label{eq:lm3}\\
&{P_{s,0}} - \beta {P_{r,\kappa}} \le 0,s \ne r\label{eq:lm4a}\\
&\ln \beta  + \alpha {\widehat{\tau} _{\max }} < 0,\label{eq:lm4}
\end{align}
where $\Psi _r^q = \frac{{\kappa\left( {{P_{r,q + 1}} - {P_{r,q}}} \right)}}{{{\widehat{\tau}_{\min }}}}$, then the system~(\ref{eq:sw}) is globally uniformly
asymptotically stable under any switching signal $\sigma(t)$ satisfying
\begin{align}
{\widehat{\tau }_{\min }} \leq {t_{k + 1}} - {t_k} \leq {\widehat{\tau}_{\max }},\forall k \in {\mathbb{N}_0}.\label{eq:emed}
\end{align}
\label{thm:sst}
\end{lem}
However, in the current form, the conditions in Lemma~\ref{thm:sst} cannot be straightforwardly applied  to our system~(\ref{eq:foon}), which is stated formally in the following proposition.
\begin{prop}
For the multi-agent system~(\ref{eq:foon}), the conditions in Lemma~\ref{thm:sst} are infeasible.\label{thm:cor}
\end{prop}
\begin{IEEEproof}
See Appendix B.
\end{IEEEproof}

To use Lemma~\ref{thm:sst}, we additionally explore whether the system~(\ref{eq:foon}) is periodic and bounded. The solutions~(\ref{eq:LP12ka})  in Lemma~\ref{thm:my0} imply that under condition~(\ref{eq:bss1}),  the state of multi-agent agent system~(\ref{eq:foon}) under fixed topology has a period $T_{r}$:
\begin{align}
T_{r} = \emph{\emph{lcm}}\left ( \frac{2\pi }{\sqrt{{\lambda_i}(\mathcal{L}_r})}; i = 2, ..., n \right ),\label{eq:pero}
\end{align}
such that
\begin{align}
\!\!\!\!\left\{ \begin{array}{l}
\hspace{-0.2cm}{{\tilde v}}\left( t \right) = -{{\tilde v}}\left( {t + \frac{T_{r}}{2}} \right),\\
\hspace{-0.2cm}{{\tilde x}}\left( t \right) = -{{\tilde x}}\left( {t \!+\! \frac{T_{r}}{2}} \right)\!, {\sigma}(t) \!=\! r \!\in\! \mathfrak{S} \hspace{0.1cm}\emph{\emph{for}}\hspace{0.1cm} t \!\in\! \left[ {{t_k},t_{k + 1}} \right).
\end{array} \right.\label{eq:ep}
\end{align}

The period $T_{\sigma(t_{k})}$ can be used to make Lemma~\ref{thm:sst} applicable to the multi-agent system~(\ref{eq:foon}) to derive a strategy on the switching times, i.e., the dwell time $\tau_{\sigma(t_{k})} = t_{k+1} - t_{k}$.
\begin{thm} Consider the second-order multi-agent system~(\ref{eq:oon}). For the given topology set $\mathfrak{S}$ satisfying~(\ref{eq:bss1}), the period $T_{\sigma(t_{k})}$ computed by~(\ref{eq:pero}), scalars $1 > \beta > 0$, $\alpha > 0$ and $\kappa \in \mathbb{N}$, if the dwell time $\tau_{\sigma(t_{k})}$ satisfy
\begin{align}
\tau_{\sigma(t_{k})} = {{\widehat \tau }_{\max }} + \mathrm{m}\frac{T_{\sigma(t_{k})}}{2}, k \in \mathbb{N}_{0}, \mathrm{m} \in \mathbb{N} \label{eq:th59}
\end{align}
where
\begin{align}
0 &< {{\widehat \tau }_{\max }} < \frac{{ - \ln \beta }}{\alpha },\label{eq:nt1} \\
0 &< {{\widehat \tau }_{\max }}   + \mathrm{m}{\frac{T_{\sigma(t_{k})}}{2}} - \left( {{\beta ^{ - \frac{1}{\kappa}}} - 1} \right)\frac{\kappa}{{\alpha  - \xi }},\label{eq:nt2}\\
\xi &< \alpha,\label{eq:ggaa1}\\
\xi &= \mathop {\max }\limits_{r \in \mathfrak{S},i = 1, \ldots ,n} \left\{ {1 - {\lambda_i}\left( {{{\cal L}_r}} \right)}, {-1 + {\lambda_i}\left( {{{\cal L}_r}} \right)} \right\},\label{eq:th10x}
\end{align}
then the second-order consensus is achieved.
\label{thm:lmr}
\end{thm}

\begin{IEEEproof}
We recall that the dynamics of fluctuations~(\ref{eq:foon})  is equivalent to~(\ref{eq:oon}). Hence, in the proof we consider only the system~(\ref{eq:foon}). We should note that the multi-agent system~(\ref{eq:foon}) can be described by a switched linear system~(\ref{eq:sw}), where
\begin{align}
\!\!\!\!\!z\left( t \right) &\triangleq \left[ {{{\tilde x}_1}\left( t \right), \ldots ,{{\tilde x}_n}\left( t \right),{{\tilde v}_1}\left( t \right), \ldots ,{{\tilde v}_n}\left( t \right)} \right]^{\top} \in {\mathbb{R}^{2n}}, \label{eq:nm0pp}\\
\!\!\!\!\!A_{\sigma(t)} &\triangleq \left[
    \begin{array}{c;{2pt/2pt}c}
        \mathbf{0} & \mathbf{I}\\ \hdashline[2pt/2pt]
        -\mathcal{L}_{\sigma(t)} & \mathbf{0}
    \end{array}
\right].\label{eq:nm0}
\end{align}

It follows from~(\ref{eq:th59})--(\ref{eq:nt2}) that the minimum and maximum dwell times defined as
\begin{align}
{\tau _{\min }} \triangleq \mathop {\min }\limits_{k \in {\mathbb{N}_0}} \left\{ {{t_{k + 1}} - {t_k}} \right\} = \mathop {\min }\limits_{k \in {\mathbb{N}_0}} \left\{ {{\tau _{\sigma \left( {{t_k}} \right)}}} \right\},\label{eq:mid}\\
{\tau _{\max }} \triangleq \mathop {\max }\limits_{k \in {\mathbb{N}_0}} \left\{ {{t_{k + 1}} - {t_k}} \right\} = \mathop {\max }\limits_{k \in {\mathbb{N}_0}} \left\{ {{\tau _{\sigma \left( {{t_k}} \right)}}} \right\},\label{eq:mad}
\end{align}
satisfy
\begin{align}
{\tau _{\max }} &< \frac{{ - \ln \beta }}{\alpha } + \mathrm{m}{\frac{T_{\sigma(t_{k})}}{2}},\label{eq:app1}\\
{\tau _{\min }} &>  {\left( {{\beta ^{ - \frac{1}{\kappa }}} - 1} \right)\frac{\kappa }{{\alpha  - \xi }}}.\label{eq:app2}
\end{align}

For each activated topology of system~(\ref{eq:foon}), let us consider the positive definite matrix
\begin{align}
P_{r,q} \triangleq \left[
    \begin{array}{c;{2pt/2pt}c}
        \hat{P}_{r,q} & \mathbf{0}\\ \hdashline[2pt/2pt]
        \mathbf{0} & \hat{P}_{r,q}
    \end{array}
\right] > 0, \label{eq:pt60}
\end{align}
where
\begin{equation}
{\hat{P}_{r,q}} \triangleq {\beta ^{ - \frac{q}{\kappa}}}h{\mathbf{I}}, \hspace{0.2cm}q = 0, \ldots ,\kappa, \forall r \in \mathfrak{S} \label{eq:pt4}
\end{equation}
with $h$ being a positive scalar. It follows from~(\ref{eq:pt4}) that
\begin{align}
&{\hat{P}_{r,q}} \triangleq {\beta ^{\frac{1}{\kappa}}}{\hat{P}_{r,q + 1}},\hspace{0.2cm}q = 0, \ldots, \kappa - 1,\forall r \in \mathfrak{S},\label{eq:pt41}\\
&{\hat{P}_{s,0}} \triangleq \beta {\hat{P}_{r,\kappa}},\hspace{0.8cm}\forall r \ne s \in \mathfrak{S}.\label{eq:pt42}
\end{align}

Substituting the matrices $P_{r,q}$~(\ref{eq:pt60}) and $A_{\sigma(t)}$~(\ref{eq:nm0}) into conditions~(\ref{eq:lm1})--(\ref{eq:lm3}) yields
\begin{align}
&R_{r,q} \buildrel \Delta \over = \left[
    \begin{array}{c;{2pt/2pt}c}
        Q_{r,q} & (\mathbf{I} - \mathcal{L}_{r})\hat{P}_{r,q}\\ \hdashline[2pt/2pt]
        (\mathbf{I} - \mathcal{L}_{r})\hat{P}_{r,q} & Q_{r,q}
    \end{array}
\right] < 0,\label{eq:pt71}\\
&\breve{R}_{r,q} \buildrel \Delta \over = \left[
    \begin{array}{c;{2pt/2pt}c}
        {\breve{Q}_{{r,q}}} & (\mathbf{I} - \mathcal{L}_{r})\hat{P}_{r,q+1}\\ \hdashline[2pt/2pt]
        (\mathbf{I} - \mathcal{L}_{r})\hat{P}_{r,q+1} & {\breve{Q}_{{r,q}}}
    \end{array}
\right] < 0,\label{eq:pt72}\\
&S_{r,\kappa} \buildrel \Delta \over = \left[
    \begin{array}{c;{2pt/2pt}c}
         - \alpha\hat{P}_{r,\kappa} & (\mathbf{I} - \mathcal{L}_{r})\hat{P}_{r,\kappa}\\ \hdashline[2pt/2pt]
        (\mathbf{I} - \mathcal{L}_{r})\hat{P}_{r,\kappa} & - \alpha\hat{P}_{r,\kappa}
    \end{array}
\right] < 0,\label{eq:pt73}
\end{align}
where
\begin{align}
{Q_{r,q}} &\triangleq \frac{\kappa}{{{\tau _{\min }}}}({{\hat P}_{r,q + 1}} - {{\hat P}_{r,q}}) - \alpha {{\hat P}_{r,q}},\label{eq:pt74}\\
{\breve{Q}_{{r,q}}} &\triangleq \frac{\kappa}{{{\tau _{\min }}}}({{\hat P}_{r,q + 1}} - {{\hat P}_{r,q}}) - \alpha {{\hat P}_{r,q + 1}}.\label{eq:pt75}
\end{align}

Let $W$ be the matrix that is orthogonal to the symmetric matrix $\mathcal{L}_{r}$, for which,
\begin{align}
{\Lambda _r} \triangleq {W^\top}{\mathcal{L}_r}W = \emph{\emph{diag}}\left\{ {0,{\lambda _2}\left( {{\mathcal{L}_r}} \right), \ldots ,{\lambda _n}\left( {{\mathcal{L}_r}} \right)} \right\}.\label{eq:ntt}
\end{align}
Then, considering the matrices $P_{r,q}$ and $\hat{P}_{r,q}$, in~(\ref{eq:pt60}) and~(\ref{eq:pt4}), the conditions~(\ref{eq:pt71})--(\ref{eq:pt73}) can be equivalently expressed in term of eigenvalue as
\begin{align}
&\frac{\kappa}{{{\tau _{\min }}}}({{\hat P}_{r,q + 1}} - {{\hat P}_{r,q}}) - \alpha {{\hat P}_{r,q}} \pm \left( {1 - {\Lambda _r}} \right){{\hat P}_{r,q}} < 0,\label{eq:pt81}\\
&\frac{\kappa}{{{\tau _{\min }}}}({{\hat P}_{r,q + 1}} \!-\! {{\hat P}_{r,q}}) \!-\! \alpha {{\hat P}_{r,q + 1}} \!\pm\! \left( {1 - {\Lambda _r}} \right){{\hat P}_{r,q+1}} < 0,\label{eq:pt82}\\
&- \alpha {{\hat P}_{r,\kappa}} \pm \left( {1 - {\Lambda _r}} \right){{\hat P}_{r,\kappa}} < 0.\label{eq:pt83}
\end{align}

In view of Lemma~\ref{thm:sst}, to prove the second-order consensus, it suffices to verify that the conditions~(\ref{eq:lm1})--(\ref{eq:lm4}) are satisfied, as carried out in the following four steps.

{\emph{Step One: }} It follows from the definitions in~(\ref{eq:pt60}),~(\ref{eq:pt41}), and~(\ref{eq:pt42}) that ${P_{s,0}} = \beta {P_{r,\kappa}},r \ne s \in \mathfrak{S}$. Thus, the condition~(\ref{eq:lm4a}) in Lemma~\ref{thm:sst} holds.

{\emph{Step Two: }} Without loss of generality, we let $\sigma \left( {{t}} \right) = r \in \mathfrak{S}$ for $t \in [t_{k}, t_{k+1})$. To obtain Lemma~\ref{thm:sst}, the considered discretized Lyapunov function for mode $r \in \mathfrak{S}$ in~\cite{xiang2014stabilization} is
\begin{align}
\hspace{-0.21cm}{V_r}\left( t \right) \!\triangleq\! \left\{ \begin{array}{l}
\hspace{-0.2cm}{z^\top}\!\!\left( t \right)P_r^{\left( q \right)}\!(\zeta)z\!\left( t \right),t \!\in\! {\mathfrak{N}_{k,q}},q \!=\! 0,1,\! \ldots,\!\kappa \!-\! 1\\
\hspace{-0.2cm}{z^\top}\!\left( t \right){P_{r,\kappa}}z\left( t \right),\hspace{0.25cm}t \in \left[ {{t_k} + {\tau _{\min }},{t_{k + 1}}} \right)
\end{array} \right.\label{eq:pt83mgmg}
\end{align}
where $P_r^{\left( q \right)}\left( \zeta  \right) \triangleq \left( {1 - \zeta } \right){P_{r,q}} + \zeta {P_{r,q + 1}}$ with $\zeta  = \frac{{\kappa \left( {t - {t_k} - {\theta _q}} \right)}}{{{\tau _{\min }}}}$, ${\mathfrak{N}_{k,q}} \triangleq \left[ {{t_k} + {\theta _q},{t_k} + {\theta _{q + 1}}} \right)$, ${\theta _{q + 1}} \triangleq \frac{{\left( {q + 1} \right){\tau _{\min }}}}{\kappa}$, ${P_{r,q}} > 0$, $q = 0,1, \ldots, \kappa - 1$.
In~\cite{xiang2014stabilization}, the purposes of the condition~(\ref{eq:lm4}) is to guarantee  that
\begin{align}
{V_{\sigma \left( {{t_k}} \right)}}\left( {{t_{k+1}}} \right) \le {\beta ^*}{V_{\sigma({t_k^ - } )}}\left( {{t_k}} \right), \label{eq:aapp22}
\end{align}
which is based on
\begin{align}
{V_{\sigma \left( {{t_k}} \right)}}\left( {{t_k} + {\widehat{\tau }_{\max }}} \right) \le {\beta ^*}{V_{\sigma({t_k^ - } )}}\left( {{t_k}} \right),\label{eq:aapp21}
\end{align}
with $1 > \beta^{*} > 0$. Noticing that dwell time relation~(\ref{eq:th59}) is equivalent to $t_{k+1} = {{t_k} + {{\widehat \tau }_{\max }} + {\rm{m}}\frac{T_{r}}{{\rm{2}}}}$, from~(\ref{eq:pt83mgmg}) we have
\begin{align}
&{V_r}\left( {{t_{k+1}}} \right) \label{eq:bbpp}\\
&= {V_r}({{t_k} + {{\widehat \tau }_{\max }} + {\rm{m}}\frac{T_{r}}{{\rm{2}}}}) \nonumber\\
&= {z^\top}({{t_k} + {{\widehat \tau }_{\max }} + {\rm{m}}\frac{T_{r}}{{\rm{2}}}}){P_{r,\kappa}}z({{t_k} + {{\widehat \tau }_{\max }} + {\rm{m}}\frac{T_{r}}{{\rm{2}}}}), \forall r \in \mathfrak{S}. \nonumber
\end{align}
Noting that ${P_{r,\kappa}} > 0$ and~(\ref{eq:ep}), we have
\begin{align}
&{z^\top}({{t_k} + {{\widehat \tau }_{\max }} + {\rm{m}}\frac{T_{r}}{{\rm{2}}}}){P_{r,\kappa}}z({{t_k} + {{\widehat \tau }_{\max }} + {\rm{m}}\frac{T_{r}}{{\rm{2}}}}) \label{eq:ccpp}\\
&={z^\top}({{t_k} + {{\widehat \tau }_{\max }}}){P_{r,\kappa}}z({{t_k} + {{\widehat \tau }_{\max }}}) \!=\! {V_r}\left( {{t_k} + {{\widehat \tau }_{\max }}} \right)\!, \forall r \!\in\! \mathfrak{S}. \nonumber
\end{align}

Combining~(\ref{eq:bbpp}) with~(\ref{eq:ccpp}) yields
\begin{align}
{V_r}\left( {{t_{k+1}}} \right) = {V_r}\left( {{t_k} + {{\widehat \tau }_{\max }}} \right)\!, \forall r \!\in\! \mathfrak{S}. \label{eq:ddpp}
\end{align}

We note that the condition~(\ref{eq:nt1}) is equivalent to $\alpha {{\widehat \tau }_{\max }} + \ln \beta  < 0$, which corresponds to the condition~(\ref{eq:lm4}) in Lemma~\ref{thm:sst}.  Then, it follows from~(\ref{eq:aapp21}) and~(\ref{eq:ddpp}) that
\begin{align}
{V_{\sigma(t_{k})}}\left( {{t_{k+1}}} \right) \!=\! {V_{\sigma \left( {{t_k}} \right)}}\left( {{t_k} \!+\! {\widehat{\tau }_{\max }}} \right) \le {\beta ^*}{V_{\sigma({t_k^ - } )}}\left( {{t_k}} \right).  \label{eq:eepp}
\end{align}
From~(\ref{eq:eepp}) and~(\ref{eq:aapp22}), we conclude that the objective of $\mathrm{m}\frac{T_{\sigma(t_{k})}}{2}, \mathrm{m} \in \mathbb{N}$, which is imposed on~(\ref{eq:th59}), is to maintain the original goal of the condition~(\ref{eq:lm4}) through keeping~(\ref{eq:eepp}) holding, while ensuring $\tau_{\max} \geq \tau_{\min}$, where $\tau_{\max}$ and $\tau_{\min}$ are given in~(\ref{eq:mid}) and~(\ref{eq:mad}), respectively.  This also means that it is the period $T_{\sigma(t_{k})}$ that makes Lemma~\ref{thm:sst} applicable to system~(\ref{eq:foon}).

{\emph{Step Three: }} Since ${{\hat P}_{r,\kappa}} > 0$, condition (\ref{eq:ggaa1}) implies $0 >  - \alpha {{\hat P}_{r,\kappa}} + \xi {{\hat P}_{r,\kappa}}$, while condition (\ref{eq:th10x}) implies $\xi  \geq  \pm \left( {1 - {\Lambda _r}} \right)$. Thus,
\begin{align}
0 >  - \alpha {{\hat P}_{r,\kappa}} + \xi {{\hat P}_{r,\kappa}} >  - \alpha {{\hat P}_{r,\kappa}} \pm \left( {1 - {\Lambda _r}} \right){{\hat P}_{r,\kappa}}.  \label{eq:att2}
\end{align}
From~(\ref{eq:pt83}), the condition~(\ref{eq:lm3}) in Lemma~\ref{thm:sst} is satisfied.

{\emph{Step Four: }} It follows from~(\ref{eq:ggaa1}) and~(\ref{eq:app2}) that
\begin{equation}
\frac{{\left( {\alpha  - \xi } \right){\tau _{\min }}}}{\kappa} + 1 > {\beta ^{ - \frac{1}{\kappa}}}.\label{eq:pt2}
\end{equation}

Considering the fact of $h > 0$, from~(\ref{eq:pt41}) and~(\ref{eq:pt2}) we have $1 + \frac{{\left( {\alpha  - \xi } \right){\tau _{\min }}}}{\kappa} > \frac{{{{\hat P}_{r,q + 1}}}}{{{{\hat P}_{r,q}}}} = {\beta ^{ - \frac{1}{\kappa}}}$, which is equivalent to
\begin{equation}
\frac{\kappa}{{{\tau _{\min }}}}({{{\hat P}_{r,q + 1}} \!-\! {{\hat P}_{r,q}}}) \!-\! \left( {\alpha  \!-\! \xi } \right){{\hat P}_{r,q}} \!<\! 0,q \!=\! 0, \ldots, \kappa \!\!-\!\! 1. \label{eq:pt5}
\end{equation}

Since $1 > \beta > 0$, relation~(\ref{eq:pt41}) implies that ${\hat{P}_{r,q}} < {\hat{P}_{r,q + 1}}$. Condition~(\ref{eq:ggaa1}) equates $\alpha - \xi> 0$. Therefore, (\ref{eq:pt5}) implies
\begin{align}
\hspace{-0.10cm}\frac{\kappa}{{{\tau _{\min }}}}({{{\hat P}_{r,q + 1}} \!\!-\!\! {{\hat P}_{r,q}}}) \!-\! \left( {\alpha \!-\! \xi} \right){{\hat P}_{r,q+1}} \!<\! 0,q \!=\! 0, \!\ldots,\! \kappa - 1.\label{eq:pt6}
\end{align}

According to~(\ref{eq:th10x}) and~(\ref{eq:att2}), we have:
\begin{align}
0 &>  - \alpha {{\hat P}_{r,q + 1}} + \xi {{\hat P}_{r,q + 1}} >  - \alpha {{\hat P}_{r,q + 1}} \pm \left( {1 - {\Lambda _r}} \right){{\hat P}_{r,q + 1}} \nonumber\\
0 &>  - \alpha {{\hat P}_{r,q}} + \xi {{\hat P}_{r,q}} >  - \alpha {{\hat P}_{r,q}} \pm \left( {1 - {\Lambda _r}} \right){{\hat P}_{r,q}},\nonumber
\end{align}which together with~(\ref{eq:pt5}) and~(\ref{eq:pt6}) imply~(\ref{eq:pt81}) and~(\ref{eq:pt82}), respectively. Thus, the conditions~(\ref{eq:lm1}) and~(\ref{eq:lm2}) in Lemma~\ref{thm:sst} hold. \end{IEEEproof}

\begin{rem} (No Zeno Behavior)
Let us first recall the formal definitions of Chattering Zeno and Genuinely Zeno behaviors~\cite{ames2005sufficient}:
\begin{itemize}
\item Chattering Zeno behavior:  $t_{k+1}- t_{k} = 0, \exists k \in  \mathbb{N}_{0}$;
\item Genuinely Zeno behavior:  $ t_{ \infty } :=  \sum_{k=0}^{\infty} (t_{k+1}- t_{k}) < \infty$ and $t_{k+1}- t_{k} > 0, \forall k \in  \mathbb{N}_{0}$.
\end{itemize}
Since $0 < \beta < 1$ and $\kappa \in \mathbb{N}$, ${{\beta ^{ - \frac{1}{\kappa}}} - 1} > 0$, which together with~(\ref{eq:ggaa1})
show that $({{\beta ^{ - \frac{1}{\kappa}}} - 1})\frac{\kappa}{{\alpha  - \xi }} > 0$. From~(\ref{eq:app2}), we conclude that
$t_{k+1}- t_{k} = \tau_{\sigma(t_{k})} \ge \tau_{\min} > 0$, $\forall k \in  \mathbb{N}_{0}$, and $t_{ \infty }$ $=$  $\sum_{k=0}^{\infty}(t_{k+1}- t_{k})$ $\ge$ $\sum_{k=0}^{\infty}\tau_{\min}$ $=$ $\lim_{k \rightarrow \infty} k\tau_{\min}$ $=$ $\infty$. Therefore, neither Chattering Zeno behavior nor Genuinely Zeno behavior can exist in the strategy on switching times presented in Theorem~\ref{thm:lmr}.
\end{rem}

\begin{rem}
Theorem~\ref{thm:lmr} shows under connected undirected communication graph, the strategy on switching times has no constraint on the magnitudes of coupling weights in achieving the asymptotic second-order consensus, which maintains the advantage of the traditional control protocol that need relative position and velocity measurements studied in~\cite{ren2007distributed}. Conditions~(\ref{eq:th59}) and~(\ref{eq:th10x}) in Theorem~\ref{thm:lmr} imply that the coupling weights affect the dwell times of switching topologies, which can further affect the convergence speed to consensus.
\end{rem}

\begin{rem}
For security, Theorem~\ref{thm:lmr} indicates that in realistic situation where the defender has no knowledge of the attack-starting time, \emph{when} the system dynamics should have changes (induced by topology switching) to detect ZDA~\cite{teixeira2012attack}, so that the changes do not destroy the system stability in the absence of attacks.
\end{rem}

\begin{figure}
\centering
\includegraphics[scale=0.42]{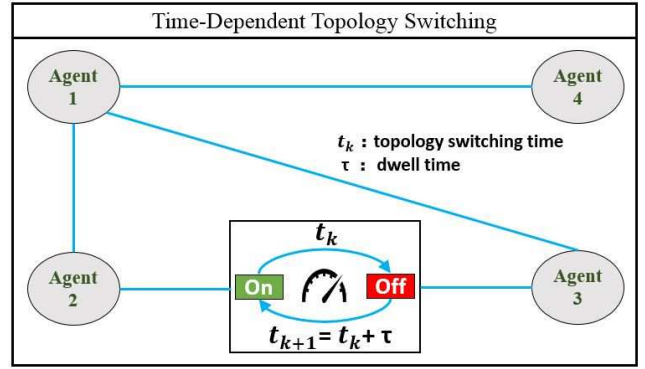}
\caption{Achieving time-dependent topology switching through controlling only one communication link: the state of controlled link $a^{\sigma(t)}_{23}$ switches between On and Off.}
\label{fig:ad1}\
\end{figure}

The strategy on switching times for asymptotic consensus is shown in Figure~\ref{fig:ad1}, which illustrates that controlling only one communication link is sufficient for the system~(\ref{eq:oon}) to switch between two different topologies, even for large-scale network.

\section{Decentralized Time-Dependent Topology Switching}
\subsection{Network Metric}
Theorem~\ref{thm:lmr} means that if the dwell time of activated topology is generated by~(\ref{eq:th59}), the asymptotic second-order consensus by Definition~\ref{thm:lmr} can be achieved through infinitely topology switching in infinite horizon, which is illustrated in Figure~\ref{fig:ad1}.   However, we should note that in real world applications of consensus algorithms, such as decentralized computation and distributed optimization, objectives must be achieved in finite time, rather than in infinite horizon, i.e., a consensus algorithm must stop \emph{when} or \emph{shortly } after the consensus error is under a preset bound $\delta > 0$ at a finite time. This motivates us to define a global network metric that can capture more properties of the multi-agent system~(\ref{eq:add}), which would tell the consensus algorithm that whether its consensus error is under its preset bound. In this context, we first define a $\delta$-consensus.
\begin{definition} ($\delta$-consensus)
Given a preset bound $\delta > 0$ and a $\mathcal{K}_{\infty}$ function $F(z(t))$ of consensus error $z(t) \triangleq [{{{\tilde x}^{\top}}}\left( t \right),{{{\tilde v}^{\top}}}\left( t \right)]^{\top}$, the $\delta$-consensus in the second-order multi-agent system~(\ref{eq:ooii}) is achieved if there exists a time $\widehat{t}_{f} < \infty$ such that $F(z(\widehat{t}_{f})) \leq \delta$.
\end{definition}

In the followings, for simplicity, we use $F(t)$ to denote $F(z(t))$. Lemma~\ref{thm:my0} implies that the system~(\ref{eq:add}) under each fixed topology can be viewed as a class of coupled oscillators. However, under some conditions, the inadmissible network metrics $F\left( t \right)$, i.e., nonzero constant scalars, of the multi-agent system~(\ref{eq:add}) are easy to exist. Take the metric $F\left( t \right) \triangleq \frac{{{{\tilde x}^\top}\left( t \right)\mathcal{L}_{r}\tilde x\left( t \right)}}{2} + \frac{{{{\tilde v}^\top}\left( t \right)\tilde v\left( t \right)}}{2}$ as an example. Differentiating it along the solutions of~(\ref{eq:add}) yields
\begin{align}
\dot F\left( t \right) &= {{\tilde v}^\top}\left( t \right)\mathcal{L}_{r}\tilde x\left( t \right) + {{\tilde v}^\top}\left( t \right)\dot{\tilde v}\left( t \right) \nonumber\\
&= {{\tilde v}^\top}\left( t \right)\mathcal{L}_{r}\tilde x\left( t \right) - {{\tilde v}^\top}\left( t \right)\mathcal{L}_{r}\tilde x\left( t \right) = 0\nonumber
\end{align}
for any $t \geq 0$, which means that for any given nonzero initial condition, the metric is a nonzero scalar over time. Therefore, this metric is inadmissible, since once a topology-switching algorithm adopts it, the topology switching would never stop, even if the $\delta$-consensus is already achieved. Obviously, such inadmissible metrics are undesirable, since they cannot capture the oscillating property of the system~(\ref{eq:add}). It is not trivial to provide a guide to construct an admissible metric. The following lemma analyzes the network metric under the strategy on switching topologies~(\ref{eq:bss}) used for detection of stealthy attack in Part-II paper~\cite{mao2017strategic}, which provides an insight for an admissible metric.
\begin{lem}
Consider the function
\begin{equation}
F\left( t \right) \triangleq \varpi\frac{{{{\tilde x}^\top}\left( t \right)\tilde x\left( t \right)}}{2} + \frac{{{{\tilde v}^\top}\left( t \right)\tilde v\left( t \right)}}{2}, \label{eq:lm2a1}
\end{equation}
with $\dot{\tilde{x}}(t) = \tilde{v}(t)  \in \mathbb{R}^{n}$. Along the solutions~(\ref{eq:LP12ka}) of the system~(\ref{eq:add}), if the Laplacian matrix $\mathcal{L}_{r}$ has distinct eigenvalues and $\varpi$ satisfies
 \begin{equation}
0 < \varpi \ne {\lambda _i}\left( {{{\mathcal{L}_{r}}}} \right), \forall i = 2, \ldots
,n,\label{eq:lm2a2}
\end{equation}
then for any nonzero initial condition and any nonzero scalar $\varphi$, the following equation never holds:
\begin{equation}
F\left( t \right) = \varphi \neq 0, \hspace{0.2cm}\emph{\emph{for}}\hspace{0.1cm}  \emph{\emph{any}} \hspace{0.1cm} t \ge 0.  \label{eq:lm2a30}
\end{equation}\label{thm:cfxo}
\end{lem}
\begin{IEEEproof}
See Appendix C.
\end{IEEEproof}

\begin{rem}
For the undirected communication network considered in this paper, there indeed exist many topologies whose associated Laplacian matrices have distinct eigenvalues. The following Lemma~\ref{thm:paa} provide a guide to design such topologies with distinct Laplacian eigenvalues:
\begin{lem}[Proposition 1.3.3 in~\cite{ada11}]
Let $\mathrm{G}$ be a connected graph with diameter $d$. Then $\mathrm{G}$ has at least $d + 1$ distinct Laplace eigenvalues.\label{thm:paa}
\end{lem}
\end{rem}

\subsection{Topology Switching Algorithm}
Recently, finite-time consensus algorithms are well developed, which can be used to estimate the global network metric precisely in finite time. Let us recall one described as follows:
\begin{lem}[Simplified Finite-Time Consensus Algorithm without External Disturbances~\cite{zuo2016distributed}]
Consider the multi-agent system
\begin{equation}
{\dot r_i} \!=\! \tilde \alpha \sum\limits_{j = 1}^n {{b_{ij}}{{\left( {{r_j} \!-\! {r_i}} \right)}^{\frac{{\bar m}}{{\bar n}}}}} \!\! + \! \tilde \beta \sum\limits_{j = 1}^n {{b_{ij}}{{\left( {{r_j} \!-\! {r_i}} \right)}^{\frac{{\bar p}}{{\bar q}}}}}, i \!=\! 1, \!\ldots\!, n\label{eq:ftc}
\end{equation}
where $\tilde{\alpha} > 0$ and $\tilde{\beta} > 0$ are the coupling strengths, $b_{ij}$ is the entry of the un-weighted adjacency matrix of an undirected connected graph and its corresponding Laplacian matrix is denoted as $\mathcal{L}_{A}$, the odd numbers $\bar{m} > 0$, $\bar{n} > 0$, $\bar{p} > 0$ and $\bar{q} > 0$ satisfy $\bar{m} > \bar{n}$ and $\bar{p} < \bar{q}$. Its global finite-time consensus can be achieved, i.e.,
\begin{equation}
{{r_i}\left( t \right) - \frac{1}{n}\sum\limits_{i = 1}^n {{r_i}\left( 0 \right)}  = 0},\hspace{0.1cm} \emph{\emph{for}} \hspace{0.1cm} t \ge S_{s},i = 1, \ldots,n\label{eq:defff}
\end{equation}
where $S_{s}$ is referred to as setting time, which means that if the running time $t$ exceeds $S_{s}$, the distributed estimation errors of the global metric $\frac{1}{n}\sum\limits_{i = 1}^n {{r_i}\left( 0 \right)}$ are zeros at $t$. Further, the setting time $S_{s}$ satisfies
\begin{equation}
S_{s} < \frac{1}{{{\lambda _2}\left( {{\mathcal{L}_A}} \right)}}\left( {\frac{{{n^{\frac{{\bar m - \bar n}}{{2\bar n}}}}}}{{\tilde \alpha }}\frac{{\bar n}}{{\bar m - \bar n}} + \frac{1}{{\tilde \beta }}\frac{{\bar q}}{{\bar q - \bar p}}} \right).\label{eq:defft}
\end{equation}\label{thm:aaa}
\end{lem}

This finite-time consensus algorithm is employed to estimate global metric to achieve $\delta$-consensus. Furthermore, from~(\ref{eq:defft}) we can see that through adjusting the control gains $\tilde{\alpha}$ and $\tilde{\beta}$, we obtain any desirable setting time $\infty > S_{s} > 0$.

Let us adjust parameters $\tilde{\alpha} > 0$ and $\tilde{\beta} > 0$ in the algorithm~(\ref{eq:ftc}), such that
\begin{equation}
\frac{1}{{{\lambda _2}\left( {{\mathcal{L}_A}} \right)}}\left( {\frac{{{n^{\frac{{\bar m - \bar n}}{{2\bar n}}}}}}{{\tilde \alpha }}\frac{{\bar n}}{{\bar m - \bar n}} + \frac{1}{{\tilde \beta }}\frac{{\bar q}}{{\bar q - \bar p}}} \right) \leq \tau_{\min}.\label{eq:mdtfop}
\end{equation}
Therefore, the setting time $S_{s}$ in~(\ref{eq:defft}) satisfies $S_{s} < \tau_{\min}$. Condition~(\ref{eq:mdtfop}) together with~(\ref{eq:defff}) and~(\ref{eq:defft}) show that if we input individual data
\begin{equation}
{F_i}\left( {{{t}_k}} \right) \triangleq \frac{\varpi }{2}\tilde x_i^2\left( {{{t}_k}} \right) + \frac{1}{2}\tilde v_i^2\left( {{{t}_k}} \right).\label{eq:idk}
\end{equation} and ${\dot{F}_i}\left( {{{t}_k}} \right)$ to the corresponding agent $i$ in the algorithm~(\ref{eq:ftc}) at time $t_{k}$, at time ${{t}_k} + \tau_{\min}$ each agent in algorithm~(\ref{eq:ftc}) will output the exact global metrics:
\begin{align}
{F_i}({{t}_k} + {{\tau _{\min }}}) &= \frac{1}{n}\sum\limits_{i = 1}^n {{F_i}({{t}_k})}  \buildrel \Delta \over = \frac{1}{n}F({{t}_k}),\label{eq:id1}\\
{{\dot F}_i}({{t}_k} + {{\tau _{\min }}}) &= \frac{1}{n}\sum\limits_{i = 1}^n {{{\dot F}_i}({{t}_k})}  \buildrel \Delta \over = \frac{1}{n}\dot F({{t}_k}).\label{eq:id2}
\end{align}

Based on Theorem~\ref{thm:lmr} and Lemma~\ref{thm:cfxo}, through employing the finite-time consensus algorithm~(\ref{eq:ftc}), we propose the decentralized time-dependent topology-switching algorithm: Algorithm~1.

\begin{algorithm}
  \caption{Decentralized  Time-Dependent Topology-Switching Algorithm}
  \KwIn{Topology set $\mathfrak{S}$ satisfying~(\ref{eq:bss}); individual functions  $F_{i}(t_{k-1})$~(\ref{eq:idk}) with $\varpi$ satisfying~(\ref{eq:lm2a2}); initial time $t_{k-1} = 0$; minimum dwell time $\tau_{\min}$ satisfying~(\ref{eq:app2});  dwell times $\tau_{r}, r \in \mathfrak{S}$, generated by~(\ref{eq:th59}); loop-stopping criteria $\delta \geq 0$.}
  \While{$F\left( t_{k-1} \right) > \delta$}
  {
  Input individuals $F_{i}({t}_{k})$ and $\dot{F}_{i}({t}_{k})$ to agent $i$ in the finite-time consensus algorithm~(\ref{eq:ftc}) at time $t_{k}$\;
  Output metrics $F({{t}_{k}})$ by~(\ref{eq:id1}) and $\dot{F}({{t}_{k}})$ by~(\ref{eq:id2}) from the finite-time consensus algorithm~(\ref{eq:ftc}) to the corresponding agents in~(\ref{eq:oon}) at time $t_{k} + \tau_{\min}$\;
    \eIf{$\dot{F}(t_{k}) = 0$ }
    {
    Switch the topology of network~(\ref{eq:oon2}) at $t_{k} +\tau_{\sigma(t_{k})}$ that satisfies:
    \begin{itemize}
     \item $\sigma(t_{k} +\tau_{\sigma(t_{k})}) \neq \sigma(t_{k})$,
     \item $\mathcal{L}_{\sigma(t_{k} +\tau)}$ has distinct eigenvalues.
     \end{itemize}
    }
    {Switch the topology of network~(\ref{eq:oon2}) at $t_{k} + \tau_{\sigma(t_{k})}$ that satisfies $\sigma(t_{k} +\tau_{\sigma(t_{k})}) \neq \sigma(t_{k})$\;}
     Update the topology-switching time: $t_{k-1} \leftarrow t_{k}$\;
     Update the metric: $F\left( t_{k-1} \right) \leftarrow F\left( t_{k} \right)$\;
     Update the topology-switching time: $t_{k} \leftarrow t_{k} + \tau_{\sigma(t_{k})}$.
     }
\end{algorithm}

\begin{thm} Consider the system~(\ref{eq:oon}). If its
topology-switching signal is generated by Algorithm~1, then the following properties hold:
\begin{description}
  \item[(i)] if the loop-stopping criteria $\delta = 0$ (in Line~1 of Algorithm 1), the agents achieve the asymptotic second-order consensus;
  \item[(ii)] if the loop-stopping criteria $\delta > 0$, the agents achieve the $\delta$-consensus through finitely topology switching, i.e., the metric $F(t)$ given in~(\ref{eq:lm2a1}) satisfies $F\left( {{t_{\bar k}}} \right) \le \delta$  with $0 < \bar k < \infty$.
\end{description}\label{thm:sstabx}
\end{thm}
\begin{IEEEproof}[Proof of Theorem~\ref{thm:sstabx}]
We first prove property (i). The loop-stopping criteria $\delta = 0$ means topology will stop switching when $F(t_{k}) = 0$. The definition of $F(t)$ in~(\ref{eq:lm2a1}) implies that $\mathop {\lim }\limits_{t \to \infty } {{F(t)  =  0}}$ is equivalent to~(\ref{eq:ooid}). This analysis means the topology switching will not stop until the asymptotic second-order consensus is achieved. Since the provided dwell time $\tau_{\sigma(t_{k})}$ in \textbf{Input} of Algorithm~1 satisfies~(\ref{eq:th59}) in Theorem~\ref{thm:lmr}, by which we conclude that property (i) holds.

We now show that property (ii) holds. Assume the function $F(t)$ is a non-zero constant over time. If $F(0) = \varphi > \delta$, from~(\ref{eq:lm2a30}) we have $F(t) = \varphi > \delta$ for any $t \geq 0$. Thus,
Line~1 of Algorithm~1 implies that in this situation the topology will never stop switching regardless of whether the $\delta$ consensus is achieved. The objective of Line 4 and Line 5 in Algorithm~1 is to switch to a topology whose associated Laplacian matrix has distinct eigenvalues when $\dot{F}(t_{k}) = 0$. Hence, by Lemma~\ref{thm:cfxo}, $F(t)$ cannot be a constant over time if $\dot{F}(t_{k}) = 0$. Therefore, by Lemma~\ref{thm:cfxo} we conclude property (ii) under Algorithm~1.
\end{IEEEproof}

\section{Simulation}
We consider a second-order system with $n = 4$ agents. Initial position and velocity conditions are randomly set as $x(0) = v(0) = {\left[ {4,2,3,4} \right]^ \top }$. We consider the topology set $\mathfrak{S}$, whose elements are shown in Figure~\ref{fig:ad1},  Table II, or  Table III, comprises only two topologies.
\begin{figure}[http]
\centering{
\begin{minipage}[b]{0.65\textwidth}
\includegraphics[width=0.8\textwidth]{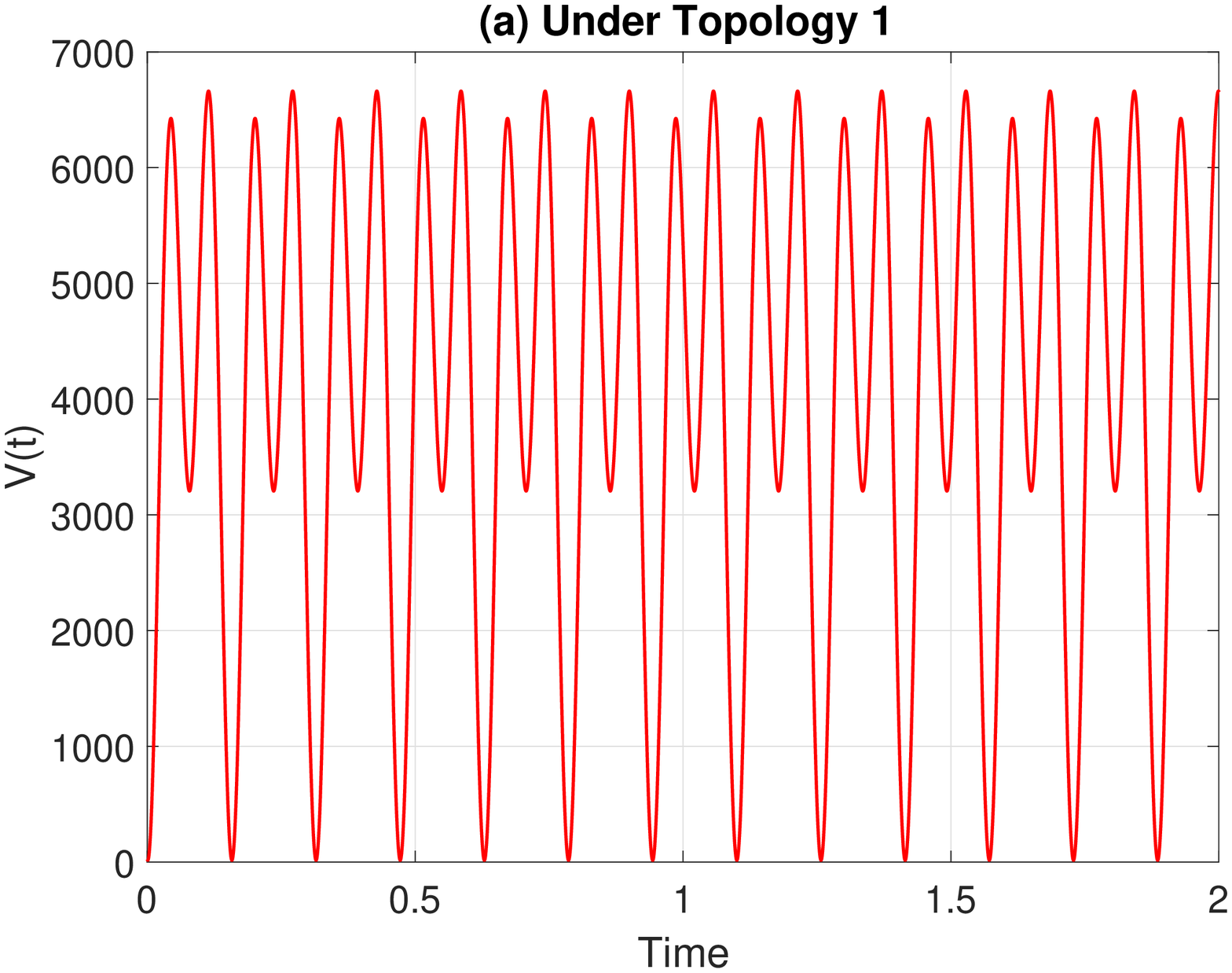} \\
\includegraphics[width=0.8\textwidth]{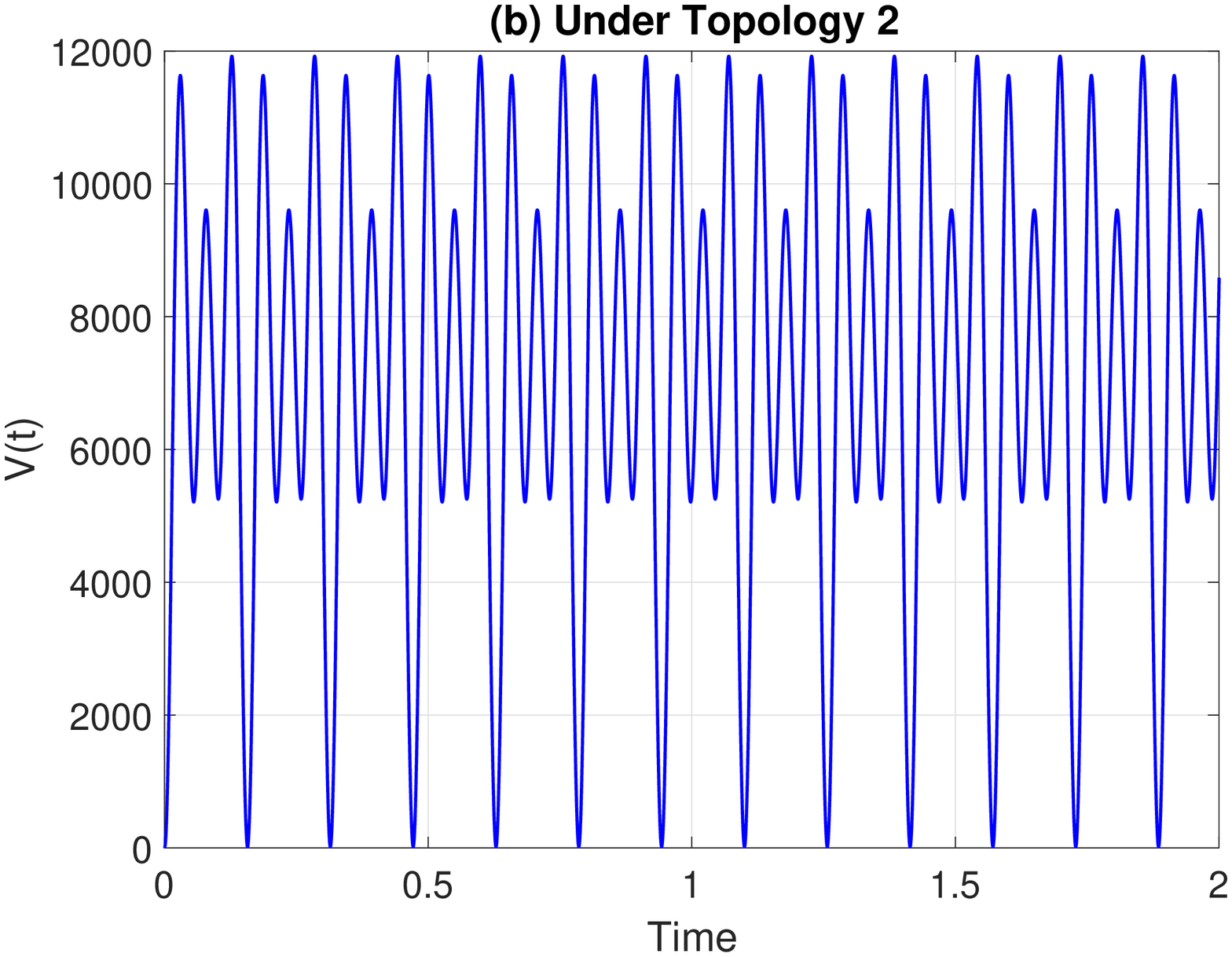}
\end{minipage}}
\caption{Trajectories of $V\left( t \right)$: 1) consensus is not achieved, 2) system is periodically oscillating.}
\label{fig:sim0}
\end{figure}

\begin{table}[http]
\caption{Topology with Large Coupling Weights}
\label{msft}
\centering
\scalebox{1}[1]{%
\begin{tabular}{  p{0.4cm}  |  p{6.5cm}}
\bottomrule
\cellcolor[gray]{0.9} ${\sigma(t)}$ & \cellcolor[gray]{0.9} $ \cellcolor[gray]{0.9} \textbf{a}^{\sigma(t)}_{12}
\hspace{0.6cm}\textbf{a}^{\sigma(t)}_{13} \hspace{0.6cm}\textbf{a}^{\sigma(t)}_{14}
\hspace{0.6cm}\textbf{a}^{\sigma(t)}_{23} \hspace{0.6cm}\textbf{a}^{\sigma(t)}_{24}
\hspace{0.6cm}\textbf{a}^{\sigma(t)}_{34}$ \\  \hline
        $1$   &\hspace{-0.1cm} 400 \hspace{0.6cm} 400 \hspace{0.6cm} 400 \hspace{0.6cm} 0 \hspace{1.2cm} 0
\hspace{0.9cm} 0 \\ \hline
        $2$   &\hspace{-0.1cm} 400 \hspace{0.60cm} 400 \hspace{0.60cm} 400 \hspace{0.6cm} 1600 \hspace{0.8cm} 0
\hspace{0.9cm} 0 \\
\bottomrule
\end{tabular}
}
\end{table}

Lemma~\ref{thm:my0} implies that the states of the system~(\ref{eq:foon}) under fixed topology keep oscillating, i.e., even with very large coupling weights, the second-order consensus cannot be achieved under fixed topology. This is numerically verified by the trajectories of $V\left( t \right)$ $=$ $\sum\limits_{i < j} {{{\left( {{x_i}\left( t \right) - {x_j}\left( t \right)} \right)}^2}}$  $+$ $\sum\limits_{i < j} {{{\left( {{v_i}\left( t \right) - {v_j}\left( t \right)} \right)}^2}}$ in Figure~\ref{fig:sim0}, where the large coupling weights are given in Table II. It verifies the topology set that includes the only two topologies in Table II satisfies~(\ref{eq:bss1}), while Figure~\ref{fig:sim0} also suggests that the states of the system~(\ref{eq:foon}) under each fixed connected topology are periodic.

\subsection{Small Coupling Weights for Asymptotic Consensus}
\begin{table}
\caption{Topology with Small Coupling Weights}
\label{msft}
\centering
\scalebox{1}[1.0]{%
\begin{tabular}{  p{0.4cm}  |  p{6.5cm}}
\bottomrule
\cellcolor[gray]{0.9} ${\sigma(t)}$ & \cellcolor[gray]{0.9} $ \cellcolor[gray]{0.9} \textbf{a}^{\sigma(t)}_{12}
\hspace{0.6cm}\textbf{a}^{\sigma(t)}_{13} \hspace{0.6cm}\textbf{a}^{\sigma(t)}_{14}
\hspace{0.6cm}\textbf{a}^{\sigma(t)}_{23} \hspace{0.6cm}\textbf{a}^{\sigma(t)}_{24}
\hspace{0.6cm}\textbf{a}^{\sigma(t)}_{34}$ \\  \hline
        $1$   &\hspace{-0.0cm} 1 \hspace{0.9cm} 1 \hspace{0.9cm} 1 \hspace{0.9cm} 0 \hspace{0.9cm} 0
\hspace{0.9cm} 0 \\ \hline
        $2$   &\hspace{-0.0cm} 1 \hspace{0.9cm} 1 \hspace{0.9cm} 1 \hspace{0.9cm} 4 \hspace{0.9cm} 0
\hspace{0.9cm} 0 \\
\bottomrule
\end{tabular}
}
\end{table}

To better show the effectiveness of Algorithm~1, we consider the same topologies but with small coupling weights, which is given in Table III. The eigenvalues of Laplacian matrices of the aforementioned topologies in Table III are computed as
\begin{align}
\!\!\!\!&\left[ {{\lambda _1}\left( {{\mathcal{L}_1}} \right)\!,{\lambda _2}\left( {{\mathcal{L}_1}} \right)\!,{\lambda _3}\left( {{\mathcal{L}_1}} \right)\!,{\lambda _4}\left( {{\mathcal{L}_1}} \right)} \right] \!=\! \left[ {0,1,1,4} \right],\label{eq:eg1}\\
\!\!\!\!&\left[ {{\lambda _1}\left( {{\mathcal{L}_2}} \right)\!,{\lambda _2}\left( {{\mathcal{L}_2}} \right)\!,{\lambda _3}\left( {{\mathcal{L}_2}} \right)\!,{\lambda _4}\left( {{\mathcal{L}_2}} \right)} \right] \!=\! \left[ {0,1,4,9} \right]. \label{eq:eg2}
\end{align}

\begin{figure}
\centering{
\begin{minipage}[b]{0.65\textwidth}
\includegraphics[width=0.8\textwidth]{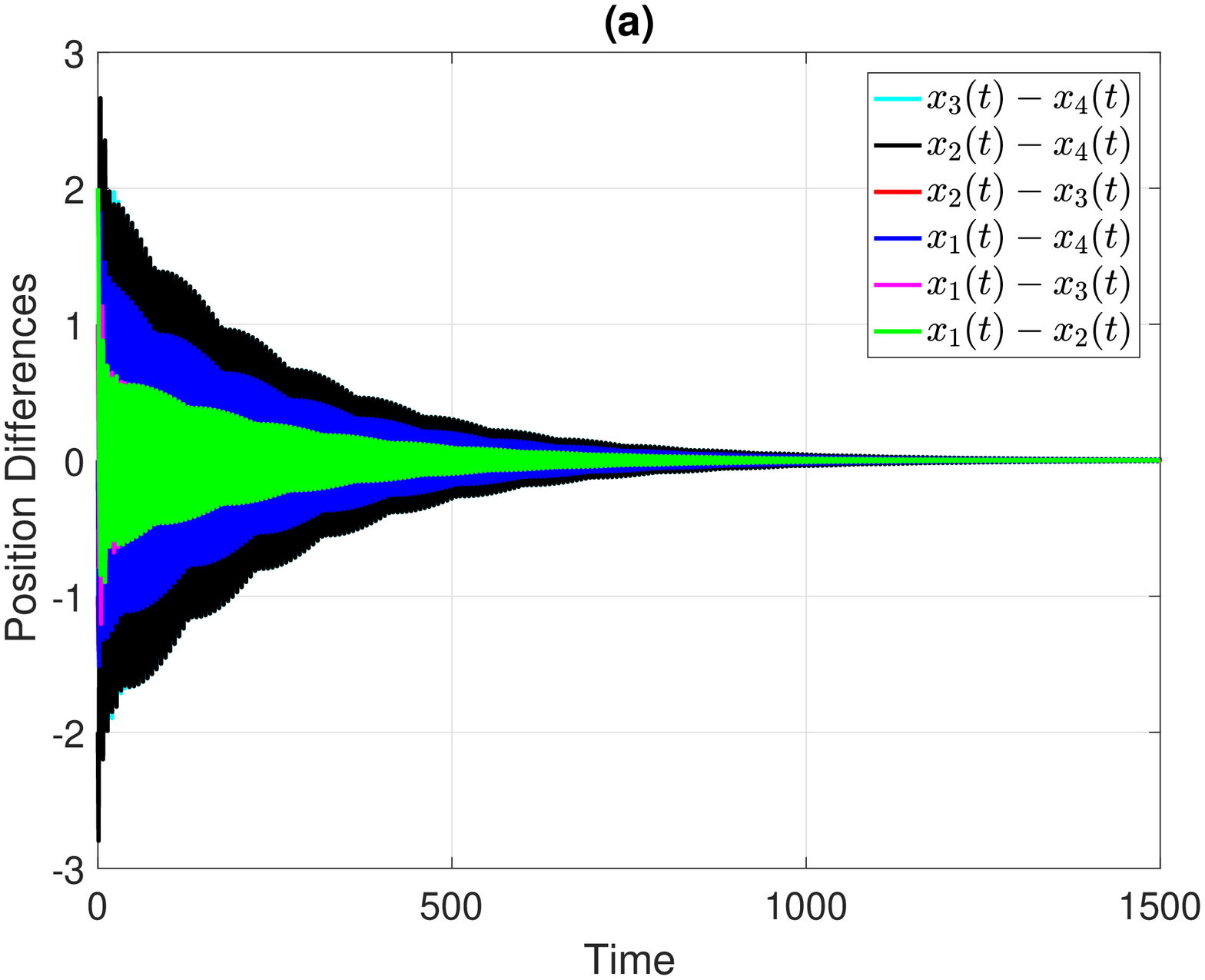} \\
\includegraphics[width=0.8\textwidth]{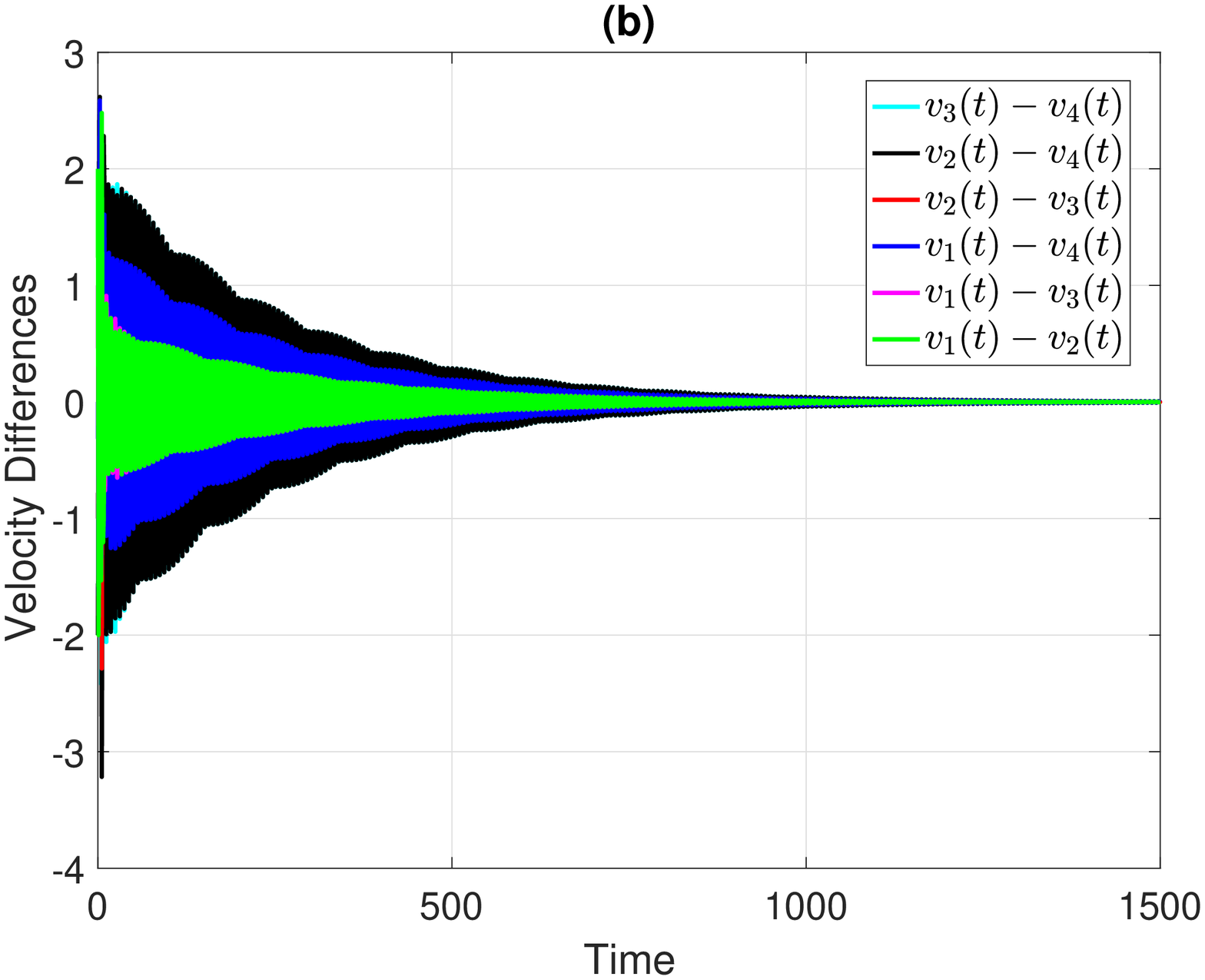}
\end{minipage}}
\caption{The trajectories of positions differences and velocities differences: the second-order consensus is achieved by Algorithm~1. }
\label{fig:sim1}
\end{figure}

It verifies the topology set that includes the only two topologies in Table III satisfies~(\ref{eq:bss1}). Using the eigenvalues computed in~(\ref{eq:eg1}) and~(\ref{eq:eg2}), by~(\ref{eq:pero}) the periods are calculated as $T_{1} = T_{2} = 2\pi$. By Theorem~\ref{thm:lmr}, we choose dwell times $\tau_{1} =  \tau_{2} = \frac{T_{1}}{2} + 0.5 =  \frac{T_{2}}{2} + 0.5$ = $\pi + 0.5$ for the two topologies 1 and 2 in Table III. Then, under Algorithm~1, trajectories of position differences and velocity differences of system~(\ref{eq:oon}) are shown in Figure~\ref{fig:sim1} (a) and Figure~\ref{fig:sim1} (b), respectively. Figure~\ref{fig:sim1} depicts that Algorithm 1 succeeds in achieving the second-order consensus. Thus, property (i) in Theorem~\ref{thm:sstabx} is numerically verified in this example.

\subsection{Finitely Topology Switching for $\delta$-Consensus}
\begin{figure}[http]
\centering{
\begin{minipage}[b]{0.65\textwidth}
\includegraphics[width=0.8\textwidth]{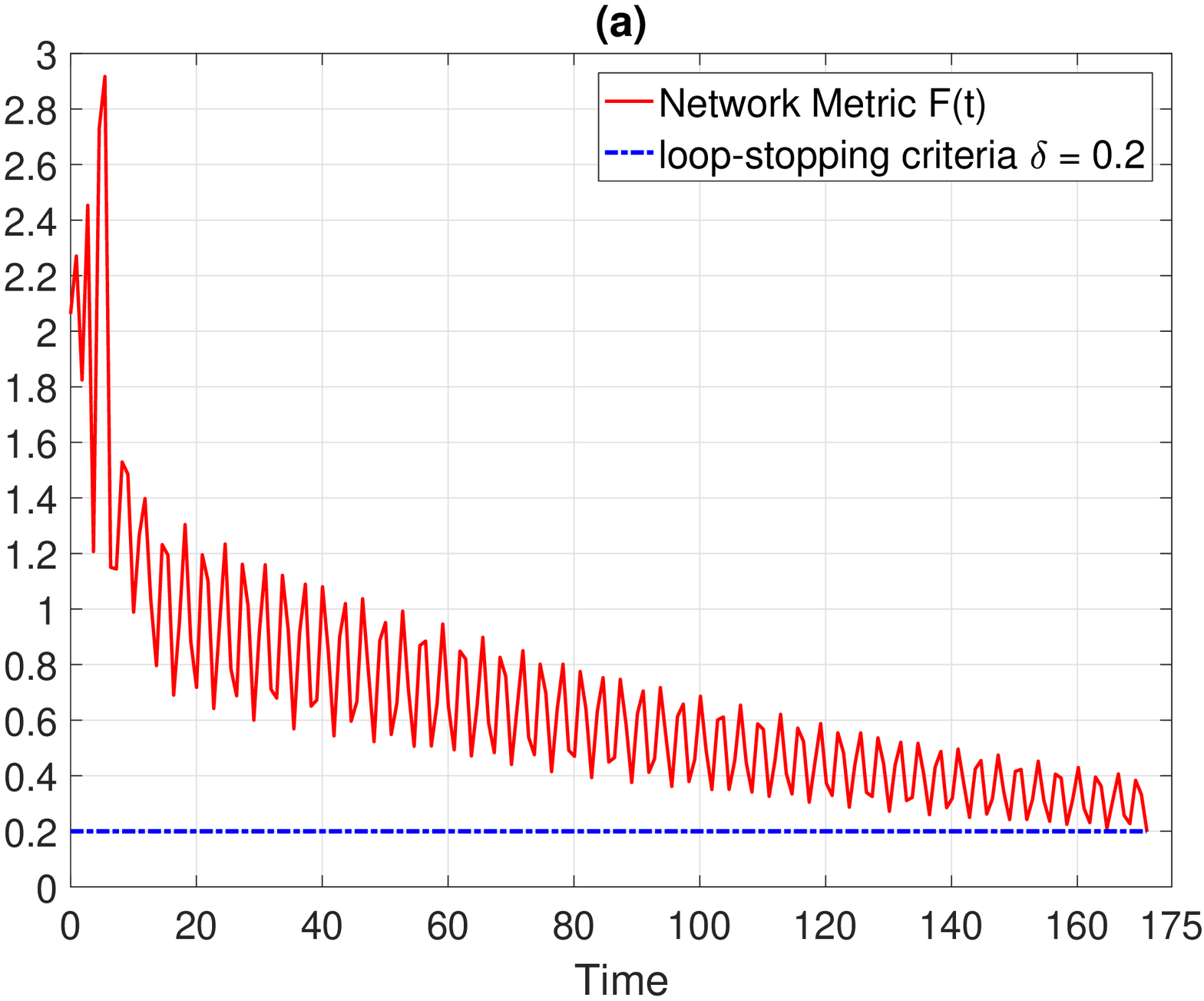} \\
\includegraphics[width=0.8\textwidth]{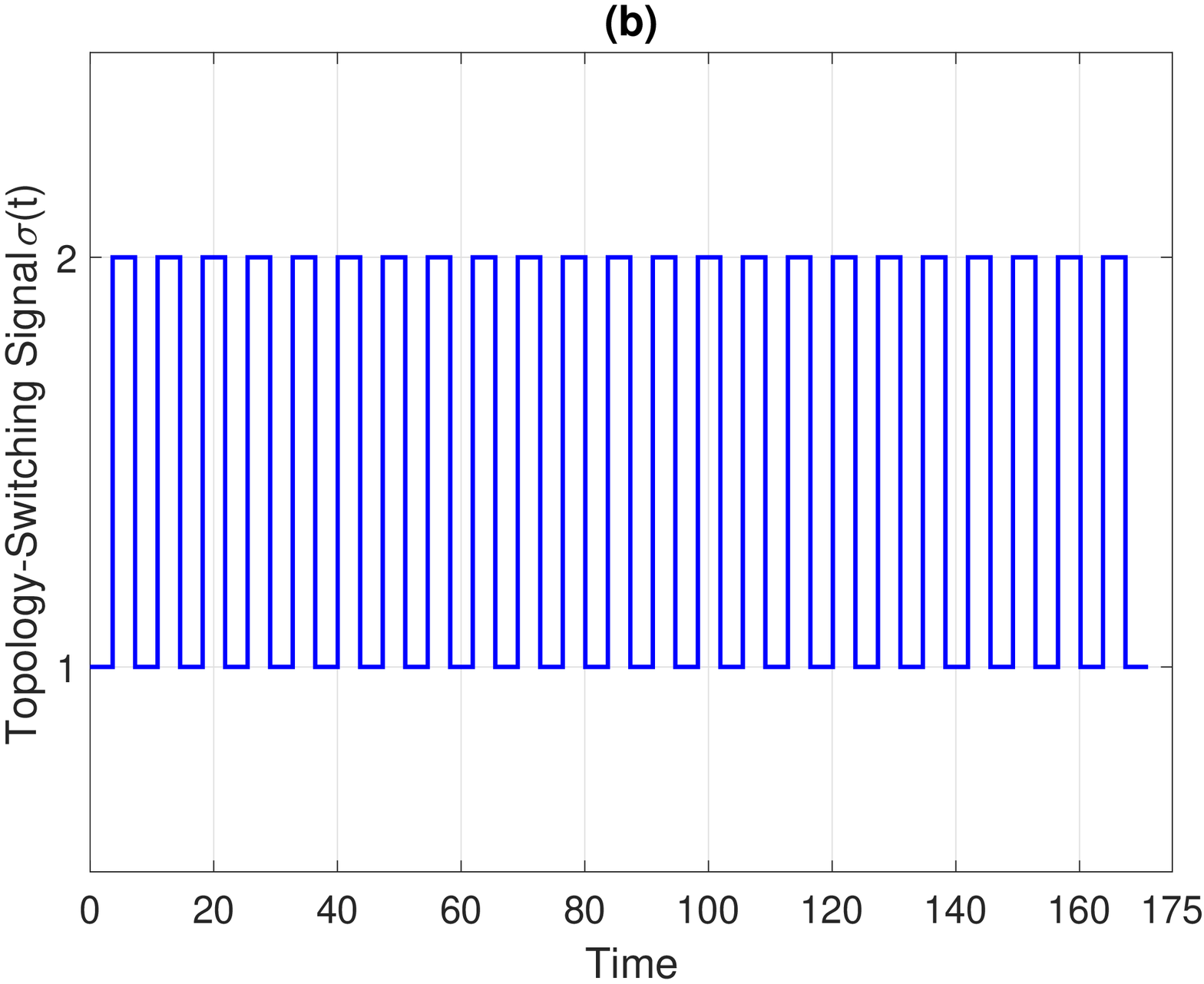}
\end{minipage}}
\caption{Trajectory of $F(t)$ and topology-switching signal $\sigma(t)$: after sixty two times topology switching, $F(t_{47}) < \delta = 0.2$.}
\label{fig:sim0x}
\end{figure}

We first note from~(\ref{eq:eg2}) that the topology 2 with its weights given in Table III has distinct Laplacian eigenvalues. Thus, Algorithm~1 can achieve the $\delta$-consensus through finitely topology switching, i.e., the property (ii) in Theorem~\ref{thm:sstabx} holds. For the condition~(\ref{eq:lm2a2}), we consider $\varpi  < \mathop {\min }\limits_{i = 2, \ldots ,n,r = 1,2} \left\{ {{\lambda _i}\left( {{{\cal L}_r}} \right)} \right\}$. From~(\ref{eq:eg2}), we choose $\varpi = 0.5$. Let the loop-stopping criteria $\delta = 0.2$. Then, under Algorithm~1, the trajectory of metric $F(t)$ given in~(\ref{eq:lm2a1}) and the topology-switching signal are shown in Figure~\ref{fig:sim0x} (a) and Figure~\ref{fig:sim0x} (b), respectively. Figure~\ref{fig:sim0x} shows that after forty-seven times topology switching, the metric $F(t)$ is under the preset error bound $\delta = 0.2$, i.e, $F(t_{47}) < \delta = 0.2$, which well verifies property (ii) in Theorem~\ref{thm:sstabx}.

\section{Conclusion}
This Part-I paper explains how to take the network topology as a control variable for the second-order multi-agent system. The obtained results highlight the merits of topology switching in achieving the second-order consensus: (i) the control protocol does not need the velocity measurements, and (ii) the topology-switching algorithm has no constraint on the magnitudes of the coupling weights. The strategy on switching times provides a basis for the strategic topology switching that is studied in Part-II paper~\cite{mao2017strategic}.

\appendices
\section{Proof of Lemma~\ref{thm:my0}}
It is well-known that the eigenvalues of Laplacian matrix of a connected graph satisfy $ {\lambda _n}(\mathcal{L}_{r}) \geq {\lambda _{n-1}}(\mathcal{L}_{r})   \ldots > {\lambda_1}(\mathcal{L}_{r}) = 0$. Since $\mathcal{L}_{r}$ is a real symmetric matrix, there exists an orthogonal matrix ${Q} \triangleq \left[ {{q_1}; \ldots; {q_n}} \right] \in {\mathbb{R}^{n \times n}}$ with ${q_i} \triangleq {\left[ {{q_{i1}},{q_{i2}}, \ldots ,{q_{in}}} \right]^\top} \in {\mathbb{R}^n}$, $i = 1$, $\ldots$, $n$, such that
\begin{align}
{Q^ \top } &= {Q^{ - 1}},\label{eq:wp1}\\
{q_{11}} &= {q_{12}} =  \ldots = {q_{1n}},\label{eq:wp2}\\
{Q^ \top }\mathcal{L}Q &= \emph{\emph{diag}}\left\{ {0,{\lambda _2}(\mathcal{L}_{r}), \ldots , {\lambda _n}(\mathcal{L}_{r})} \right\} \buildrel \Delta
\over = \Lambda.\label{eq:wp3}
\end{align}

We denote ${\rm X}\left( s \right) \triangleq \mathfrak{L}\left\{ {\tilde{x} \left( t \right)} \right\}$, where $\mathfrak{L}(\cdot)$ stands for the Laplace transform operator. The Laplace transform
of the dynamics~(\ref{eq:add}) can be obtained as
\begin{equation}
s{\rm{X}}\left( s \right) - \tilde x\left( 0 \right) =  - \frac{\mathcal{L}_{r}}{s}{\rm{X}}\left( s \right) +
\frac{{\tilde v\left( 0 \right)}}{s},\label{eq:LP3}
\end{equation}
which is equivalent to
\begin{equation}
{\rm{X}}\left( s \right) = ({{s^2}{\textbf{I}} + \mathcal{L}_{r}})^{-1}\left( {s\tilde x\left( 0 \right) +
\tilde v\left( 0 \right)} \right).\label{eq:LP4}
\end{equation}

We let $\tilde \Lambda(s)  \triangleq \emph{\emph{diag}}\left\{ {\frac{1}{s},\frac{s}{{{s^2} + {\lambda _2}(\mathcal{L}_{r})}}, \ldots ,\frac{s}{{{s^2} + {\lambda _n}(\mathcal{L}_{r})}}} \right\}$ and $\bar \Lambda(s)  \triangleq \emph{\emph{diag}}\left\{ {\frac{1}{{{s^2}}},\frac{1}{{{s^2} + {\lambda _2}(\mathcal{L}_{r})}}, \ldots ,\frac{1}{{{s^2} + {\lambda _n}(\mathcal{L}_{r})}}} \right\}$. It follows from~(\ref{eq:wp1}) and (\ref{eq:wp3}) that
\begin{align}
\! \! \! \! &({{s^2}{\textbf{I}} + {\cal L}})^{-1}{s{\textbf{I}}}  \! =\!
\left({Q\left( {{s^2}{\textbf{I}} + \Lambda } \right){Q^\top}}\right)^{-1}{s{\textbf{I}}} \! =\!  Q\tilde \Lambda(s) {Q^\top},\label{eq:a1}\\
\! \! \! \! &({{s^2}{\textbf{I}} + {\cal L}})^{-1}{{\textbf{I}}}  =
\left({Q\left( {{s^2}{\textbf{I}} + \Lambda } \right){Q^\top}}\right)^{-1}{{\textbf{I}}} = Q\bar \Lambda(s) {Q^\top}.\label{eq:a2}
\end{align}
We let ${{\rm X}_i}\left( s \right)$, $i = 1, \ldots, n$, be the $i^{\emph{\emph{th}}}$ entry of ${{\rm X}}\left( s \right)$.
From~(\ref{eq:LP4})--(\ref{eq:a2}),
\begin{equation}
\begin{array}{l}
{{\rm{X}}_i}\left( s \right) = {q_{1i}}q_1^ \top \left( {\frac{1}{s}\tilde x\left( 0 \right) +
\frac{1}{{{s^2}}}\tilde v\left( 0 \right)} \right)\\
\hspace{1.35cm} + \sum\limits_{l = 2}^n {\frac{s}{{{s^2} + {\lambda _l}(\mathcal{L}_{r})}}} {q_{li}}q_l^ \top \left( {\tilde
x\left( 0 \right) + \frac{1}{s}\tilde v\left( 0 \right)} \right).
\end{array}\label{eq:LP6}
\end{equation}
It follows from~(\ref{eq:usef1}),~(\ref{eq:usef1no}) and~(\ref{eq:wp2}) that
\begin{align}
q_1^ \top \tilde x\left( 0 \right) &= 0,\label{eq:LP9}\\
q_1^ \top \tilde v\left( 0 \right) &= 0.\label{eq:LP9no}
\end{align}

Substituting~(\ref{eq:LP9}) and~(\ref{eq:LP9no}) into~(\ref{eq:LP6}) yields
\begin{align}
{{\rm{X}}_i}\left( s \right)
\!  =\!  \sum\limits_{l = 2}^n \! {\frac{s}{{{s^2}  \! +\!   {\lambda _l}(\mathcal{L}_{r})}}} {q_{li}}q_l^\top \! \tilde x\! \left( 0 \right)  \! +\!
\sum\limits_{l = 2}^n \! {\frac{1}{{{s^2}  \! + \!  {\lambda _l}(\mathcal{L}_{r})}}} {q_{li}}q_l^\top \! \tilde v\! \left( 0 \right).\nonumber
\end{align}
then the solution~(\ref{eq:LP12ka})  can be obtained
immediately from the inverse Laplace transform of ${{\rm{X}}_i}\left( s \right)$.

\section{Proof of Proposition~\ref{thm:cor}}
We prove this proposition via a contradiction. We  assume the conditions in Lemma~\ref{thm:sst} are feasible along the solutions of the dynamics~(\ref{eq:foon}). We write the multi-agent system~(\ref{eq:foon}) in the form of switched system~(\ref{eq:sw}), where ${A_{\sigma (t)}}$ is given by~(\ref{eq:nm0}). We next consider a positive definite matrix: \\
${P_{r,q}} \triangleq \left[ {\begin{array}{*{20}{c}}
{{G_{r,q}}}&{{V_{r,q}}}\\
{{V^{\top}_{r,q}}}&{{S_{r,q}}}
\end{array}} \right] > 0$, where ${G_{r,q}},{V_{r,q}},{S_{r,q}} \in {\mathbb{R}^{n \times n}}$. Then, from~(\ref{eq:nm0}), we have \begin{align}
\Gamma_{r,q} &\buildrel \Delta \over =  A_r^ \top {P_{r,q}} + {P_{r,q}}{A_r} \nonumber\\
&= \left[ {\begin{array}{*{20}{c}}
\hspace{-0.2cm}{ - {{\cal L}_r}{V^{\top}_{r,q}} - {V_{r,q}}{{\cal L}_r}}&{{G_{r,q}} - {S_{r,q}}{\mathcal{L}_r}}\\
\hspace{-0.2cm}{{G_{r,q}} - {\mathcal{L}_r}{S_{r,q}}}&{{V_{r,q}+V^{\top}_{r,q}}}
\end{array}} \right] .\label{eq:adn0}
\end{align}

For a connected undirected graph, ${\cal L}_r \geq 0$. From~(\ref{eq:adn0}), we observe that if $V_{r,q}+V^{\top}_{r,q}$ $\geq 0$ ($\leq 0$), $- {\mathcal{L}_r}V_{r,q}^ \top  - {V_{r,q}}{\mathcal{L}_r}$ $\leq 0$ ($\geq 0$). Thus, $\Gamma_{r,q}$ has at least one non-negative diagonal entry. If $V_{r,q}+V^{\top}_{r,q}$ is indefinite, let us assume that all of the diagonal entries of $\Gamma_{r,q}$ are negative, $V_{r,q}+V^{\top}_{r,q}$ would have at least one positive eigenvalue; then, it is straightforward to verify that $- {\mathcal{L}_r}V_{r,q}^ \top  - {V_{r,q}}{\mathcal{L}_r}$ has at least one non-negative diagonal entry, which contradicts with the assumption that ``all of the diagonal entries of $\Gamma_{r,q}$ are negative." Therefore, we conclude that $\Gamma_{r,q}$ has at least one non-negative diagonal entry.

Without loss of generality, we let $[\Gamma_{r,q}]_{1,1} \geq  0$. Let us denote ${p_{r,q}} \triangleq {\left[ {{P_{r,q}}} \right]_{1,1}}$. Considering ${\Psi}_r^q = \frac{{\kappa \left( {{{P}_{r,q + 1}} - {{P}_{r,q}}} \right)}}{{{\widehat{\tau }_{\min }}}}$ (given in Lemma~\ref{thm:sst}) and~(\ref{eq:adn0}), then it follows from~(\ref{eq:lm1}) and~(\ref{eq:lm4a}) that
\begin{align}
&\frac{\kappa}{{{\widehat{\tau }_{\min }}}}\left( {{{p}_{r,q + 1}} \!-\! {{p}_{r,q}}} \right) \!<\! \alpha {{p}_{r,q}}, \forall r \in \mathfrak{S}, q \!=\! 0, 1, \ldots, \kappa \!-\! 1\label{eq:ladlm1}\\
&{{p}_{s,0}} \leq \beta {{p}_{r,\kappa}}, \hspace{2.1cm}\forall s \ne r \in \mathfrak{S}.\label{eq:ladlm3}
\end{align}

Since ${{P}_{r,q}} > 0$ for $\forall r \in \mathfrak{S}$ and $\forall q = 0, 1, \ldots, \kappa$, ${{p}_{r,q}} > 0$ for $\forall r \in \mathfrak{S}$ and $\forall q = 0, 1, \ldots, \kappa$. Thus, (\ref{eq:ladlm1}) is equivalent to
\begin{equation}
{\widehat{\tau }_{\min }} > \frac{\kappa}{\alpha }\left( {\frac{{{{p}_{r,q + 1}}}}{{{{p}_{r,q}}}} - 1} \right),\forall q = 0, \ldots ,\kappa-1,\forall r \in \mathfrak{S}.\label{eq:ccl2}
\end{equation}

Condition~(\ref{eq:lm4}) is equivalent to ${\widehat{\tau }_{\max }} < \frac{{ - \ln \beta }}{\alpha}$ that, together with~(\ref{eq:ccl2}) and the fact of ${\widehat{\tau }_{\max }} \ge {\widehat{\tau }_{\min }}$, yields
\begin{equation}
- \ln \beta  \!>\! \kappa\left( {\frac{{{{p}_{r,q + 1}}}}{{{{p}_{r,q}}}} - 1} \right),\forall q \!=\! 0, \!\ldots\!,\kappa-1,\forall r \in \mathfrak{S}.\label{eq:ccl3}
\end{equation}

Noting that (\ref{eq:ladlm3}) implies $\frac{{{{p}_{r,\kappa}}}}{{{{p}_{s,0}}}} \ge \frac{1}{\beta } > 1$ and $\frac{{{{p}_{s,\kappa}}}}{{{\breve{p}_{r,0}}}} \ge \frac{1}{\beta } > 1$, we have
\begin{align}
&{p}_{s,0}^{ - 1}{p}_{r,0}^{ - 1}{{p}_{s,\kappa}}{{p}_{r,\kappa}}\label{eq:ccl5}\\
&= \prod\limits_{q = 0}^{\kappa - 1} {{p}_{s,q}^{ - 1}{p}_{r,q}^{ - 1}{{p}_{s,q + 1}}{{p}_{r,q + 1}}}  \ge {\beta ^{ - 2}} > 1
, \forall r \ne s \in \mathfrak{S}. \nonumber
\end{align}

Considering~(\ref{eq:ccl5}), we pick up a number $\hat{q} \in \{0, 1,\ldots, \kappa -1\}$ such that ${p}_{s,\hat q}^{ - 1}{p}_{r,\hat q}^{ - 1}{{p}_{s,\hat q + 1}}{{p}_{r,\hat q + 1}} \ge {\beta ^{ - \frac{2}{\kappa}}} > 1,\forall r \ne s \in \mathfrak{S}$, which also implies that under one of the indices $s$ and $r$, say $r$, that:
\begin{equation}
{p}_{r,\hat q}^{ - 1}{{p}_{r,\hat q + 1}} \ge {\beta ^{ - \frac{1}{\kappa}}} > 1,r \in \mathfrak{S}.\label{eq:ccl6}
\end{equation}

Combining~(\ref{eq:ccl3}) with~(\ref{eq:ccl6}) yields $- \ln \beta  > \kappa({{\beta ^{ - \frac{1}{\kappa}}} - 1})$ which is equivalent to
\begin{equation}
\beta  < {e^{\kappa({1 - {\beta ^{\frac{{ - 1}}{\kappa}}}})}}, \beta \in (0,1).\label{eq:ccl7}
\end{equation}
For~(\ref{eq:ccl7}), let us consider the function
\begin{equation}
\tilde{g}\left( \beta  \right) \triangleq {e^{\kappa(1 - {\beta ^{\frac{{ - 1}}{\kappa}}})}} - \beta, \beta \in (0,1).\label{eq:ccl7a}
\end{equation}
It is straightforward to verify from~(\ref{eq:ccl7a}) that $\tilde{g}(0) = 0$ and $\tilde{g}(1) = 0$, which imply that under any fixed $\kappa \in \mathbb{N} < \infty$, the function $\tilde{g}\left( \beta  \right)$ has at least one extreme point over the interval $(0,1)$ and $\tilde{g}\left( \beta  \right) > 0$ might happen in its interval $(0,1)$.

\begin{figure}[http]
\centering
\includegraphics[scale=0.44]{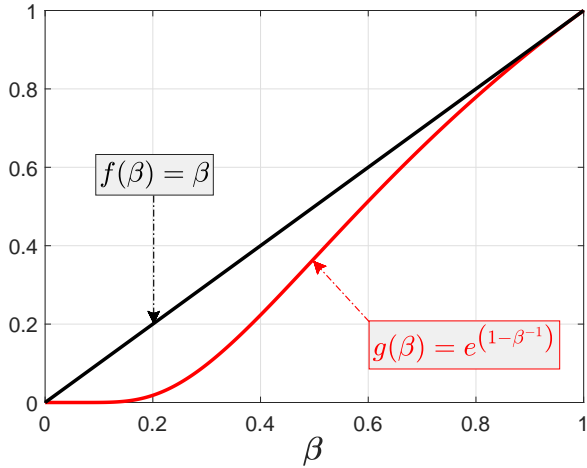}
\caption{Take $\kappa = 1$ as an example, ${e^{\left( {1 - {\beta ^{ - 1}}} \right)}} < \beta$ for any $\beta \in (0,1)$.}
\label{fig:mid}
\end{figure}

The derivative of $\tilde{g}\left( \beta  \right)$~(\ref{eq:ccl7a}) with respect to $\beta \in (0,1)$  is obtained as
\begin{equation}
\frac{{\partial \tilde{g}\left( \beta  \right)}}{{\partial \beta }} = {e^{\kappa(1 - {\beta ^{\frac{{ - 1}}{\kappa}}})}}{\beta ^{ - ({\frac{1}{\kappa} + 1})}} - 1.\label{eq:ccl7vb}
\end{equation}
For $\frac{{\partial \tilde{g}\left( \beta_{*}  \right)}}{{\partial \beta_{*} }} = 0$, we obtain from~(\ref{eq:ccl7vb}):
 ${e^{\kappa(1 - {\beta_{*}^{\frac{{ - 1}}{\kappa}}})}} = {\beta_{*}^{({\frac{1}{\kappa} + 1})}}$, substituting which into~(\ref{eq:ccl7a}) yields that at extreme point: $\tilde{g}\left( {{\beta _ * }} \right) = \beta _*^{({\frac{1}{\kappa} + 1} )} - {\beta _ * } = {\beta _ * }({\beta _*^{\frac{1}{\kappa}} - 1}) < 0$. Then, noting $\tilde{g}(0) = 0$ and $\tilde{g}(1) = 0$, we conclude that under any fixed $\kappa \in \mathbb{N}$, $\tilde{g}(\beta) < 0$ for any $\beta \in (0,1)$. Therefore, (\ref{eq:ccl7}) never holds (an illustration is given in Figure~\ref{fig:mid}, where we take $\kappa= 1$ as an example), which is a contradiction, and this completes the proof.

\section{Proof of Lemma~\ref{thm:cfxo}}
We first reproduce the following well-known result.
\begin{lem}~\cite{hom1991topics}
The determinant of the Vandermonde matrix
\begin{align}
\mathrm{H} \triangleq \left[ {\begin{array}{*{20}{c}}
1&1& \cdots &1\\
{{a_1}}&{{a_2}}& \cdots &{{a_n}}\\
{a_1^2}&{a_2^2}& \cdots &{a_n^2}\\
{a_1^3}&{a_2^3}& \cdots &{a_n^3}\\
 \vdots & \vdots & \vdots & \vdots \\
{a_1^{n - 1}}&{a_2^{n - 1}}& \cdots &{a_n^{n - 1}}
\end{array}} \right] \in \mathbb{R}^{n \times n}, \nonumber
\end{align}
is $\det \left( {\rm{H}} \right)  = {\left( { - 1} \right)^{\frac{{{n^2} - n}}{2}}}\prod\limits_{i < j} {\left( {{a_i} - {a_j}} \right)}.$
\label{thm:amm}
\end{lem}

We prove Lemma~\ref{thm:cfxo} via a contradiction. We first assume that~(\ref{eq:lm2a30}) holds, from which we obtain
\begin{equation}
\frac{{{\partial ^\mathrm{d}}}}{{\partial t^{\mathrm{d}}}}F\left( t \right) = 0, \hspace{0.2cm} \emph{\emph{for}} \hspace{0.1cm}t \ge 0, \forall \mathrm{d} \in \mathbb{N}.  \label{eq:l31}
\end{equation}
It follows from the dynamics~(\ref{eq:add}) that
\begin{align}
\ddot{\tilde x}\left( t \right)  &= -  \mathcal{L}_{r}\tilde{x}\left( t \right),\hspace{0.2cm} \emph{\emph{for}} \hspace{0.2cm}t \ge 0 \label{eq:l34}\\
\ddot{\tilde{v}}\left( t \right) &= - \mathcal{L}_{r} \tilde{v}\left( t \right),\hspace{0.2cm} \emph{\emph{for}} \hspace{0.2cm}t \ge 0.\label{eq:l35}
\end{align}
The rest of the proof is divided into two steps.

{\emph{Step One}}: For the dynamics~(\ref{eq:add}), relations~(\ref{eq:l31})--(\ref{eq:l35}) imply
\begin{align}
&\frac{{{\partial ^{\rm{d}}}}}{{\partial {t^{\rm{d}}}}}F\left( t \right)\nonumber\\
& = \frac{{{\partial ^{{\rm{d - 1}}}}}}{{\partial {t^{{\rm{d - 1}}}}}}\dot F\left( t \right)\nonumber\\
& =  \frac{{{\partial ^{{\rm{d - 1}}}}}}{{\partial {t^{{\rm{d - 1}}}}}}{{\tilde v}^\top }\!\!\left( t \right)\left( {\varpi \textbf{I} - \mathcal{L}_{r}} \right)\tilde x\left( t \right)\nonumber\\
 & = \frac{{{\partial^{{\rm{d - 2}}}}}}{{\partial {t^{{\rm{d - 2}}}}}}2\left( {{{\tilde v}^ \top }\!\!\left( t \right)\left( {\varpi {\bf{I}} \!-\! {\cal L}_{r}} \right)\tilde v\left( t \right) \!-\! {{\tilde x}^ \top }\left( t \right){\cal L}_{r}\left( {\varpi {\bf{I}} \!-\! {\cal L}_{r}} \right)\tilde x\left( t \right)} \right)\nonumber\\
  &= \!\!\frac{{{\partial ^{{\rm{d - 4}}}}}}{{\partial {t^{{\rm{d - 4}}}}}}\!-\!{2^3}\!\!\left( {{{\tilde v}^\top }\!\!\left( t \right)\!{\cal L}_{r}\!\left( {\varpi {\bf{I}} \!-\! {\cal L}_{r}} \right)\!\tilde v\!\left( t \right) \!-\! {{\tilde x}^\top }\!\!\left( t \right)\!{\mathcal{L}_{r}^2}\!\left( {\varpi {\bf{I}} \!-\! {\cal L}_{r}} \right)\!\tilde x\!\left( t \right)} \right) \nonumber\\
  &= \frac{{{\partial ^{{\rm{d - 2m}}}}}}{{\partial {t^{{\rm{d - 2m}}}}}}\left( {{{( - 1)}^{{\rm{m}} - 1}}{2^{2{\rm{m}} - 1}}} \right)\left( {{{\tilde v}^ \top }\!\!\left( t \right){{\cal L}_{r}^{{\rm{m}} - 1}}\left( {\varpi {\bf{I}} - {\cal L}_{r}} \right)\tilde v\left( t \right)} \right.\nonumber\\
  &\hspace{4.5cm}\left. { - {{\tilde x}^ \top }\!\!\left( t \right){{\cal L}_{r}^{\rm{m}}}\left( {\varpi {\bf{I}} - {\cal L}_{r}} \right)\tilde x\left( t \right)} \right) \nonumber\\
 & = 0, \hspace{0.1cm} \emph{\emph{for}} \hspace{0.1cm}t \ge 0, \forall \mathrm{m} \in \mathbb{N}, \forall \mathrm{d} > 2\mathrm{m} \in \mathbb{N} \nonumber
\end{align}
which then implies
\begin{align}
&{{\tilde x}^\top}\!\!\left( t \right){\mathcal{L}_{r}^\mathrm{m}}\left( {\varpi \textbf{I} - \mathcal{L}_{r}} \right)\tilde x\left( t \right) \nonumber\\
&= {{\tilde v}^\top}\!\!\left( t \right){{\mathcal{L}_{r}}^{\mathrm{m} - 1}}\left( {\varpi \textbf{I} -  \mathcal{L}_{r}} \right)\tilde v\left( t \right), \hspace{0.1cm} \emph{\emph{for}} \hspace{0.1cm}t \ge 0, \forall\mathrm{m} \in \mathbb{N}. \label{eq:l32}
\end{align}
It follows from~(\ref{eq:l32}),~(\ref{eq:wp1}), and~(\ref{eq:wp3}) that
\begin{align}
&{{\tilde x}^\top}\!\!\left( t \right)Q{\Lambda ^\mathrm{m}}\left( {\varpi \textbf{I} - \Lambda } \right){Q^ \top }\tilde x\left( t \right) \nonumber\\
&= {{\tilde v}^ \top }\!\!\left( t \right)Q{\Lambda ^{\mathrm{m} \!-\! 1}}\left( {\varpi \textbf{I} \!-\! \Lambda } \right){Q^ \top }\tilde v\left( t \right),\hspace{0.1cm} \emph{\emph{for}} \hspace{0.1cm}t \ge 0, \forall \mathrm{m} \!\in\! \mathbb{N} \label{eq:l33x}
\end{align}
where $\Lambda$ is given in~(\ref{eq:wp3}), and ${Q} = \left[ {{q_1}; \ldots; {q_n}} \right] \in {\mathbb{R}^{n \times n}}$ is an orthogonal matrix produced from the real symmetric matrix $\mathcal{L}_{r}$,  where $q_{1}, q_{2}, \ldots, q_{n}$ are orthogonal vectors that correspond to the eigenvalues $0 = {\lambda _1}(\mathcal{L}_{r}), \lambda_{2}(\mathcal{L}_{r}), \ldots, {\lambda
_n}(\mathcal{L}_{r})$. Let us define:
\begin{align}
\widehat x\left( t \right) \triangleq {Q^ \top }\tilde x\left( t \right),\label{eq:lx1}\\
\widehat v\left( t \right) \triangleq {Q^ \top }\tilde v\left( t \right).\label{eq:lx2}
\end{align}

Considering~(\ref{eq:LP9}) and~(\ref{eq:LP9no}), from~(\ref{eq:lx1}) and~(\ref{eq:lx2}), we have $\widehat x_{1}\left( t \right) = 0$ and $\widehat v_{1}\left( t \right) = 0$, respectively. Hence, $\widehat x\left( t \right)$ and $\widehat v\left( t \right)$ can be rewritten as
\begin{align}
\widehat x\left( t \right) &= {\left[ {0,{{\widehat x}_2}\left( t \right),{{\widehat x}_3}\left( t \right), \ldots ,{{\widehat x}_n}\left( t \right)} \right]^\top} \in {\mathbb{R}^n},\label{eq:mlx1}\\
\widehat v\left( t \right) &= {\left[ {0,{{\widehat v}_2}\left( t \right),{{\widehat v}_3}\left( t \right), \ldots ,{{\widehat v}_n}\left( t \right)} \right]^\top} \in {\mathbb{R}^n}.\label{eq:mlx2}
\end{align}

Noting the matrix $\Lambda$ given in~(\ref{eq:wp3}), we conclude that the relation~(\ref{eq:l33x}) is equivalent to
\begin{align}
&\sum\limits_{i = 2}^n {\lambda _i^\mathrm{m}(\mathcal{L}_{r})\left( {\varpi - {\lambda _i}(\mathcal{L}_{r})} \right)\widehat x_i^2\left( t \right)} \nonumber\\
&= \sum\limits_{i = 2}^n {\lambda_i^{\mathrm{m} - 1}(\mathcal{L}_{r})\left( {\varpi - {\lambda _i}(\mathcal{L}_{r})} \right)\widehat v_i^2\left( t \right)}, \hspace{0.1cm} \emph{\emph{for}} \hspace{0.1cm}t \ge 0, \forall \mathrm{m} \in \mathbb{N},\nonumber
\end{align}
which is also equivalent to
\begin{align}
&\sum\limits_{i = 2}^n {\lambda _i^{{\rm{m}} - 1}(\mathcal{L}_{r})\!\left( {\varpi  - {\lambda _i}(\mathcal{L}_{r})} \right)\left( {{\lambda _i}(\mathcal{L}_{r})\widehat x_i^2\left( t \right) - \widehat v_i^2\!\left( t \right)} \right)}\nonumber\\
&= 0, \forall \mathrm{m} \in \mathbb{N}, t \ge 0.\label{eq:l33}
\end{align}

Let us define:
\begin{align}
\!\!&{{\widehat z}_i}\left( t \right) \triangleq \left( {\varpi  - {\lambda _i}(\mathcal{L}_{r})} \right)\left( {{\lambda _i}(\mathcal{L}_{r})\widehat x_i^2\left( t \right) - \widehat v_i^2\left( t \right)} \right),\label{eq:lx1x}\\
\!\!&{\rm H} \triangleq \left[ {\begin{array}{*{20}{c}}
1& \ldots &1&1\\
{{\lambda _2}(\mathcal{L}_{r})}& \ldots &{{\lambda _{n - 1}}(\mathcal{L}_{r})}&{{\lambda _n}(\mathcal{L}_{r})}\\
{\lambda _2^2(\mathcal{L}_{r})}& \ldots &{\lambda _{n - 1}^2(\mathcal{L}_{r})}&{\lambda _n^2(\mathcal{L}_{r})}\\
 \vdots & \vdots & \vdots & \vdots \\
{\lambda _2^{n - 2}(\mathcal{L}_{r})}& \ldots & {\lambda _{n - 1}^{n - 2}(\mathcal{L}_{r})} & {\lambda _n^{n - 2}(\mathcal{L}_{r})}
\end{array}} \right].\label{eq:lx2x}
\end{align}

Using (\ref{eq:lx1x}) and~(\ref{eq:lx2x}) in conjunction with (\ref{eq:l33}), we obtain
\begin{equation}
{\rm H}\widehat z\left( t \right) = {\mathbf{0}_{n - 1}},\label{eq:l33opv}
\end{equation}
where $\widehat z\left( t \right) \triangleq {\left[ {{{\widehat z}_2}\left( t \right), \ldots ,{{\widehat z}_n}\left( t \right)} \right]^\top} \in {\mathbb{R}^{n - 1}}$.

The condition that $\mathcal{L}_{r}$  has distinct eigenvalues implies that the elements $\lambda_{l}(\mathcal{L}_{r})$, $l = 2$, $\ldots$, $n$, in the Vandermonde matrix ${\rm H} \in \mathbb{R}^{(n-1) \times (n-1)}$ given by~(\ref{eq:lx2x}) are also distinct. Hence, by Lemma~\ref{thm:amm}, $\det \left( {\rm H} \right) \neq 0$. Therefore, we conclude that the solution of~(\ref{eq:l33opv}) is $\widehat z\left( t \right) = {{\bf{0}}_{n - 1}}$. Furthermore, considering the condition~(\ref{eq:lm2a2}), we obtain from~(\ref{eq:lx1x}) that ${\lambda _i}(\mathcal{L}_{r})\widehat x_i^2\left( t \right) = \widehat v_i^2\left( t \right), \forall i = 2, \ldots, n$, for $t \geq 0$, which is equivalent to
\begin{equation}
\widehat v\left( t \right) = \Delta \widehat x\left( t \right), \hspace{0.1cm} \emph{\emph{for}} \hspace{0.1cm}t \ge 0\label{eq:l33op}
\end{equation}
where
\begin{equation}
\Delta  \triangleq \emph{\emph{diag}}\left\{0, { \pm \sqrt {{\lambda _2}(\mathcal{L}_{r})} , \ldots , \pm \sqrt {{\lambda _n}(\mathcal{L}_{r})} } \right\} \in {\mathbb{R}^{{n} \times  {n} }}.\label{eq:la1}
\end{equation}

{\emph{Step Two}}: In view of the dynamics~(\ref{eq:add}), it follows from~(\ref{eq:lm2a1}),~(\ref{eq:wp3}),~(\ref{eq:lx1}) and~(\ref{eq:lx2}) that
\begin{align}
\frac{{{\partial ^\mathrm{m}}}}{{\partial {t^\mathrm{m}}}}F\left( t \right) &= \frac{{{\partial ^{{\rm{m - 1}}}}}}{{\partial {t^{{\rm{m - 1}}}}}}\dot F\left( t \right) \nonumber\\
&= \frac{{{\partial ^{{\rm{m - 1}}}}}}{{\partial {t^{{\rm{m - 1}}}}}}\left( {{{\tilde v}^\top}\left( t \right)\tilde x\left( t \right) + {{\tilde v}^\top}\left( t \right)\dot{\tilde v}\left( t \right)} \right)\nonumber\\
& = \frac{{{\partial ^{{\rm{m - 1}}}}}}{{\partial {t^{{\rm{m - 1}}}}}}\left( {{{\tilde v}^\top}\left( t \right)\left( {\varpi \mathbf{I} - \mathcal{L}_{r}} \right)\tilde x\left( t \right)} \right)\nonumber\\
& = \frac{{{\partial ^{{\rm{m - 1}}}}}}{{\partial {t^{{\rm{m - 1}}}}}}\left( {{{\tilde v}^\top}\left( t \right)Q\left( {\varpi \mathbf{I} - \Lambda } \right){Q^\top}\tilde x\left( t \right)} \right)\nonumber\\
& = \frac{{{\partial ^{{\rm{m - 1}}}}}}{{\partial {t^{{\rm{m - 1}}}}}}\left( {{{\widehat v}^\top}\left( t \right)\left( {\varpi \mathbf{I} - \Lambda } \right)\widehat x\left( t \right)} \right). \nonumber
\end{align}
Then, under~(\ref{eq:l31}), considering~(\ref{eq:l33op}),~(\ref{eq:l34}) and~(\ref{eq:l35}) we obtain
\begin{align}
\frac{{{\partial ^\mathrm{m}}}}{{\partial {t^\mathrm{m}}}}F\left( t \right)  &= \frac{{{\partial ^{{\rm{m - 1}}}}}}{{\partial {t^{{\rm{m - 1}}}}}}\left( {{{\widehat x}^\top}\left( t \right)\Delta \left( {\varpi \mathbf{I} - \Lambda } \right)\widehat x\left( t \right)} \right)\nonumber\\
 &= \frac{{{\partial ^{{\rm{m - 2}}}}}}{{\partial {t^{{\rm{m - 2}}}}}}\left( {2{{\widehat v}^\top}\left( t \right)\Delta \left( {\varpi \mathbf{I} - \Lambda } \right)\widehat x\left( t \right)} \right)\nonumber\\
 &= \frac{{{\partial ^{{\rm{m - 2}}}}}}{{\partial {t^{{\rm{m - 2}}}}}}\left( {2{{\widehat x}^\top}\left( t \right){\Delta ^2}\left( {\varpi \mathbf{I} - \Lambda } \right)\widehat x\left( t \right)} \right)\nonumber\\
 &= {2^{\mathrm{m} - 1}}{{\widehat x}^\top}\left( t \right){\Delta ^\mathrm{m}}\left( {\varpi \mathbf{I} - \Lambda } \right)\widehat x\left( t \right)\nonumber\\
 &= 0,\hspace{0.1cm} \emph{\emph{for}} \hspace{0.1cm}t \ge 0, \forall \mathrm{m} \in \mathbb{N} \nonumber
\end{align}
which can be expressed equivalently in term of the entries of the matrices $\Lambda$ in~(\ref{eq:wp3}) and $\Delta$ in~(\ref{eq:la1}) as
\begin{equation}
\sum\limits_{i = 2}^n { {{\left( \pm {{\lambda _i}(\mathcal{L}_{r})} \right)}^{\frac{\mathrm{m}}{2}}}} \left( {\omega  - {\lambda _i}(\mathcal{L}_{r})} \right)\widehat x_i^2\left( t \right) = 0, \hspace{0.05cm} \emph{\emph{for}} \hspace{0.05cm}t \ge 0, \forall \mathrm{m} \in \mathbb{N}.\nonumber
\end{equation}

Let $\mathrm{m}$ belong to the set of even numbers. Noting the defined Vandermonde matrix ${\rm H}$ in~(\ref{eq:lx2x}),  from the above equality we have
\begin{equation}
{\rm H}\breve{\Lambda} \bar z\left( t \right) = {\mathbf{0}_{n - 1}}, \hspace{0.1cm} \emph{\emph{for}} \hspace{0.1cm}t \ge 0.\label{eq:la1x01}
\end{equation}
where $\breve{\Lambda}  \triangleq \emph{\emph{diag}}\left\{ {{\lambda _2}(\mathcal{L}_{r}), \ldots ,{\lambda _n}(\mathcal{L}_{r})} \right\} \in {\mathbb{R}^{\left( {n - 1} \right) \times \left( {n - 1} \right)}}$ and $\bar z\left( t \right) \triangleq {\left[ {{{\bar z}_2}\left( t \right), \ldots ,{{\bar z}_n}\left( t \right)} \right]^\top} \in {\mathbb{R}^{n - 1}}$ with
\begin{equation}
{{\bar z}_i}\left( t \right) \triangleq \left( {\varpi  - {\lambda _i}(\mathcal{L}_{r})} \right)\widehat x_i^2\left( t \right),i = 2, \ldots ,n.\label{eq:la1x02}
\end{equation}

As obtained in the previous step ($\emph{Step One}$) of the proof, $\det \left( {\rm H} \right) \neq 0$. Since $\det \left( {\breve{\Lambda} } \right) \ne 0$, we have  ${\rm{det}}\left( {{\rm{H}}\breve{\Lambda} } \right) = \det \left( {\rm{H}} \right)\det \left( {\breve{\Lambda} } \right) \ne 0$. Therefore, the solution of~(\ref{eq:la1x01}) is $\bar z\left( t \right) = {\mathbf{0}_{n - 1}}$, for $t \ge 0.$ We note that the condition~(\ref{eq:lm2a2}) is equivalent to $\varpi  - {\lambda _i}(\mathcal{L}_{r}) \ne 0,\forall i = 2, \ldots ,n,$ which in conjunction with~(\ref{eq:la1x02}), implies that the obtained solution $\bar z\left( t \right) = {\mathbf{0}_{n - 1}}$, for $t \ge 0$, is equivalent to $\widehat x_i^2\left( t \right) = 0$, for $t \geq 0$, $\forall i = 2, \ldots, n$. Now, considering~(\ref{eq:l33op}), we have $\widehat v_i^2\left( t \right) = 0$, for $t \geq 0$, $\forall i = 2, \ldots, n$. Finally, noting the orthogonal matrix $Q$ is full-rank, from~(\ref{eq:lx1})--(\ref{eq:mlx2}) we conclude that $\tilde x\left( t \right) = {\mathbf{0}_n}$ and $\tilde v\left( t \right) = {\mathbf{0}_n}$, for $t \ge 0$, which contradicts with~(\ref{eq:lm2a30}).

\ifCLASSOPTIONcaptionsoff
  \newpage
\fi

\section*{Acknowledgment}
We thank Dr. Sadegh Bolouki, Dr. Hamidreza Jafarnejadsani, and Dr. Pan Zhao for valuable discussions.

\bibliographystyle{IEEEtran}

\bibliography{ref}

\begin{thebibliography}{10}
\providecommand{\url}[1]{#1}
\csname url@samestyle\endcsname
\providecommand{\newblock}{\relax}
\providecommand{\bibinfo}[2]{#2}
\providecommand{\BIBentrySTDinterwordspacing}{\spaceskip=0pt\relax}
\providecommand{\BIBentryALTinterwordstretchfactor}{4}
\providecommand{\BIBentryALTinterwordspacing}{\spaceskip=\fontdimen2\font plus
\BIBentryALTinterwordstretchfactor\fontdimen3\font minus
  \fontdimen4\font\relax}
\providecommand{\BIBforeignlanguage}[2]{{%
\expandafter\ifx\csname l@#1\endcsname\relax
\typeout{** WARNING: IEEEtran.bst: No hyphenation pattern has been}%
\typeout{** loaded for the language `#1'. Using the pattern for}%
\typeout{** the default language instead.}%
\else
\language=\csname l@#1\endcsname
\fi
#2}}
\providecommand{\BIBdecl}{\relax}
\BIBdecl

\bibitem{cdccs18}
Y.~Mao, E.~Akyol, and Z.~Zhang, ``Second-order consensus for multi-agent
  systems by time-dependent topology switching,'' in \emph{Decision and Control
  (CDC), 2018 IEEE 57th Conference on}.\hskip 1em plus 0.5em minus 0.4em\relax
  IEEE, 2018, pp. 6151--6156.

\bibitem{mao2017strategic}
------, ``Strategic topology switching for security-{Part II}: detection \&
  switching topologies,'' \emph{https://arxiv.org/abs/1711.11181}.

\bibitem{olfati2004consensus}
R.~Olfati-Saber and R.~M. Murray, ``Consensus problems in networks of agents
  with switching topology and time-delays,'' \emph{IEEE Transactions on
  automatic control}, vol.~49, no.~9, pp. 1520--1533, 2004.

\bibitem{jadbabaie2003coordination}
A.~Jadbabaie, J.~Lin, and A.~S. Morse, ``Coordination of groups of mobile
  autonomous agents using nearest neighbor rules,'' \emph{IEEE Transactions on
  automatic control}, vol.~48, no.~6, pp. 988--1001, 2003.

\bibitem{fax2004information}
J.~A. Fax and R.~M. Murray, ``Information flow and cooperative control of
  vehicle formations,'' \emph{IEEE transactions on automatic control}, vol.~49,
  no.~9, pp. 1465--1476, 2004.

\bibitem{ren2007distributed}
W.~Ren and E.~Atkins, ``Distributed multi-vehicle coordinated control via local
  information exchange,'' \emph{International Journal of Robust and Nonlinear
  Control}, vol.~17, no. 10-11, pp. 1002--1033, 2007.

\bibitem{tsitsiklis1984problems}
J.~N. Tsitsiklis, ``Problems in decentralized decision making and
  computation,'' Massachusetts Inst. of Tech. Cambridge Lab for Information and
  Decision Systems, Tech. Rep., 1984.

\bibitem{nedic2009distributed}
A.~Nedi$\acute{\emph{\emph{c}}}$ and A.~Ozdaglar, ``Distributed subgradient
  methods for multi-agent optimization,'' \emph{IEEE Transactions on Automatic
  Control}, vol.~54, no.~1, pp. 48--61, 2009.

\bibitem{lu2015consensus}
L.~Y. Lu and C.~C. Chu, ``Consensus-based droop control synthesis for multiple
  dics in isolated micro-grids,'' \emph{IEEE Transactions on Power Systems},
  vol.~30, no.~5, pp. 2243--2256, 2015.

\bibitem{li2006global}
Q.~Li and D.~Rus, ``Global clock synchronization in sensor networks,''
  \emph{IEEE Transactions on computers}, vol.~55, no.~2, pp. 214--226, 2006.

\bibitem{abdessameud2009attitude}
A.~Abdessameud and A.~Tayebi, ``Attitude synchronization of a group of
  spacecraft without velocity measurements,'' \emph{IEEE Transactions on
  Automatic Control}, vol.~54, no.~11, pp. 2642--2648, 2009.

\bibitem{chung2009cooperative}
S.~J. Chung and J.~J.~E. Slotine, ``Cooperative robot control and concurrent
  synchronization of lagrangian systems,'' \emph{IEEE Transactions on
  Robotics}, vol.~25, no.~3, pp. 686--700, 2009.

\bibitem{johnson2014synchronization}
B.~B. Johnson, S.~V. Dhople, A.~O. Hamadeh, and P.~T. Krein, ``Synchronization
  of nonlinear oscillators in an lti electrical power network,'' \emph{IEEE
  Transactions on Circuits and Systems I: Regular Papers}, vol.~61, no.~3, pp.
  834--844, 2014.

\bibitem{olfati2006flocking}
R.~Olfati-Saber, ``Flocking for multi-agent dynamic systems: Algorithms and
  theory,'' \emph{IEEE Transactions on automatic control}, vol.~51, no.~3, pp.
  401--420, 2006.

\bibitem{gazi2004stability}
V.~Gazi and K.~M. Passino, ``Stability analysis of social foraging swarms,''
  \emph{IEEE Transactions on Systems, Man, and Cybernetics, Part B
  (Cybernetics)}, vol.~34, no.~1, pp. 539--557, 2004.

\bibitem{tanner2007flocking}
H.~G. Tanner, A.~Jadbabaie, and G.~J. Pappas, ``Flocking in fixed and switching
  networks,'' \emph{IEEE Transactions on Automatic control}, vol.~52, no.~5,
  pp. 863--868, 2007.

\bibitem{lawton2003decentralized}
J.~R. Lawton, R.~W. Beard, and B.~J. Young, ``A decentralized approach to
  formation maneuvers,'' \emph{IEEE transactions on robotics and automation},
  vol.~19, no.~6, pp. 933--941, 2003.

\bibitem{ren2004decentralized}
W.~Ren and R.~Beard, ``Decentralized scheme for spacecraft formation flying via
  the virtual structure approach,'' \emph{Journal of Guidance, Control, and
  Dynamics}, vol.~27, no.~1, pp. 73--82, 2004.

\bibitem{rahili2017distributed}
S.~Rahili and W.~Ren, ``Distributed continuous-time convex optimization with
  time-varying cost functions,'' \emph{IEEE Transactions on Automatic Control},
  vol.~62, no.~4, pp. 1590--1605, 2017.

\bibitem{yu2010some}
W.~Yu, G.~Chen, and M.~Cao, ``Some necessary and sufficient conditions for
  second-order consensus in multi-agent dynamical systems,'' \emph{Automatica},
  vol.~46, no.~6, pp. 1089--1095, 2010.

\bibitem{mei2016distributed}
J.~Mei, W.~Ren, and J.~Chen, ``Distributed consensus of second-order
  multi-agent systems with heterogeneous unknown inertias and control gains
  under a directed graph,'' \emph{IEEE Transactions on Automatic Control},
  vol.~61, no.~8, pp. 2019--2034, 2016.

\bibitem{qin2016leaderless}
J.~Qin, C.~Yu, and B.~D. Anderson, ``On leaderless and leader-following
  consensus for interacting clusters of second-order multi-agent systems,''
  \emph{Automatica}, vol.~74, pp. 214--221, 2016.

\bibitem{ai2016second}
X.~Ai, S.~Song, and K.~You, ``Second-order consensus of multi-agent systems
  under limited interaction ranges,'' \emph{Automatica}, vol.~68, pp. 329--333,
  2016.

\bibitem{yu2011second}
W.~Yu, W.~X. Zheng, G.~Chen, W.~Ren, and J.~Cao, ``Second-order consensus in
  multi-agent dynamical systems with sampled position data,''
  \emph{Automatica}, vol.~47, no.~7, pp. 1496--1503, 2011.

\bibitem{huang2016some}
N.~Huang, Z.~Duan, and G.~R. Chen, ``Some necessary and sufficient conditions
  for consensus of second-order multi-agent systems with sampled position
  data,'' \emph{Automatica}, vol.~63, pp. 148--155, 2016.

\bibitem{abdessameud2010consensus}
A.~Abdessameud and A.~Tayebi, ``On consensus algorithms for double-integrator
  dynamics without velocity measurements and with input constraints,''
  \emph{Systems \& Control Letters}, vol.~59, no.~12, pp. 812--821, 2010.

\bibitem{xia2018event}
H.~Xia, W.~X. Zheng, and J.~Shao, ``Event-triggered containment control for
  second-order multi-agent systems with sampled position data,'' \emph{ISA
  transactions}, vol.~73, pp. 91--99, 2018.

\bibitem{Song17}
Q.~Song, F.~Liu, G.~Wen, J.~Cao, and X.~Yang, ``Distributed position-based
  consensus of second-order multiagent systems with continuous/intermittent
  communication,'' \emph{IEEE transactions on cybernetics}, vol.~47, no.~8, pp.
  1860--1871, 2017.

\bibitem{ren2005consensus}
W.~Ren and R.~W. Beard, ``Consensus seeking in multiagent systems under
  dynamically changing interaction topologies,'' \emph{IEEE Transactions on
  automatic control}, vol.~50, no.~5, pp. 655--661, 2005.

\bibitem{psillakis2017consensus}
H.~E. Psillakis, ``Consensus in networks of agents with unknown high-frequency
  gain signs and switching topology,'' \emph{IEEE Transactions on Automatic
  Control}, vol.~62, no.~8, pp. 3993--3998, 2017.

\bibitem{lin2010consensus}
P.~Lin and Y.~Jia, ``Consensus of a class of second-order multi-agent systems
  with time-delay and jointly-connected topologies,'' \emph{IEEE Transactions
  on Automatic Control}, vol.~55, no.~3, pp. 778--784, 2010.

\bibitem{su2016distributed}
S.~Su and Z.~Lin, ``Distributed consensus control of multi-agent systems with
  higher order agent dynamics and dynamically changing directed interaction
  topologies,'' \emph{IEEE Transactions on Automatic Control}, vol.~61, no.~2,
  pp. 515--519, 2016.

\bibitem{xie2005consensus}
G.~Xie and L.~Wang, ``Consensus control for networks of dynamic agents via
  active switching topology,'' in \emph{International Conference on Natural
  Computation}.\hskip 1em plus 0.5em minus 0.4em\relax Springer, 2005, pp.
  424--433.

\bibitem{M18}
Y.~Mao and Z.~Zhang, ``Second-order consensus for multi-agent systems by
  state-dependent topology switching,'' in \emph{American Control Conference
  (ACC), 2018}.\hskip 1em plus 0.5em minus 0.4em\relax IEEE, 2018, pp.
  3392--3397.

\bibitem{mao2018synchronization}
Y.~Mao and E.~Akyol, ``Synchronization of coupled harmonic oscillators by
  time-dependent topology switching,'' \emph{IFAC-PapersOnLine}, vol.~51,
  no.~23, pp. 402--407, 2018.

\bibitem{allerton18}
------, ``Detectability of cooperative zero-dynamics attack,'' in \emph{2018
  56th Annual Allerton Conference on Communication, Control, and Computing
  (Allerton)}, 2018, pp. 227--234.

\bibitem{1582237}
A.~D. Ames, A.~Abate, and S.~Sastry, ``Sufficient conditions for the existence
  of zeno behavior,'' in \emph{Decision and Control, 2005 44th IEEE Conference
  on}.\hskip 1em plus 0.5em minus 0.4em\relax IEEE, 2005, pp. 696--701.

\bibitem{1470118}
A.~D. Ames and S.~Sastry, ``Characterization of zeno behavior in hybrid systems
  using homological methods,'' in \emph{2005 Annual American Control Conference
  (ACC)}.\hskip 1em plus 0.5em minus 0.4em\relax IEEE, 2005, pp. 1160--1165.

\bibitem{teixeira2012attack}
A.~Teixeira, D.~P{\'e}rez, H.~Sandberg, and K.~H. Johansson, ``Attack models
  and scenarios for networked control systems,'' in \emph{Proceedings of the
  1st international conference on High Confidence Networked Systems}.\hskip 1em
  plus 0.5em minus 0.4em\relax ACM, 2012, pp. 55--64.

\bibitem{xiang2014stabilization}
W.~Xiang and J.~Xiao, ``Stabilization of switched continuous-time systems with
  all modes unstable via dwell time switching,'' \emph{Automatica}, vol.~50,
  no.~3, pp. 940--945, 2014.

\bibitem{ames2005sufficient}
A.~Ames, A.~Abate, and S.~Sastry, ``Sufficient conditions for the existence of
  zeno behavior,'' in \emph{Decision and Control, 2005 and 2005 European
  Control Conference. CDC-ECC'05. 44th IEEE Conference on}.\hskip 1em plus
  0.5em minus 0.4em\relax IEEE, 2005, pp. 696--701.

\bibitem{ada11}
A.~E. Brouwer and W.~H. Haemers, \emph{Spectra of graphs}.\hskip 1em plus 0.5em
  minus 0.4em\relax Springer Science \& Business Media, 2011.

\bibitem{zuo2016distributed}
Z.~Zuo and L.~Tie, ``Distributed robust finite-time nonlinear consensus
  protocols for multi-agent systems,'' \emph{International Journal of Systems
  Science}, vol.~47, no.~6, pp. 1366--1375, 2016.

\bibitem{hom1991topics}
R.~A. Hom and C.~R. Johnson, ``Topics in matrix analysis,'' \emph{Cambridge UP,
  New York}, 1991.

\end{thebibliography}

\end{document}